\documentclass[fleqn,usenatbib]{mnras}

\usepackage[T1]{fontenc}

\DeclareRobustCommand{\VAN}[3]{#2}
\let\VANthebibliography\thebibliography
\def\thebibliography{\DeclareRobustCommand{\VAN}[3]{##3}\VANthebibliography}

\usepackage{graphicx}	
\usepackage{amsmath}	
\usepackage{amssymb}	
\usepackage{mathtools}
\usepackage{cleveref}

\newcommand{\feh}{\left [ \mathrm{Fe}/\mathrm{H} \right ]}

\title[Magellanic Mayhem]{Magellanic Mayhem: Metallicities and Motions}

\author[J. Grady et al.]{
J. Grady,$^{1}$\thanks{E-mail: jrg71,vasily,nwe@cam.ac.uk}
V. Belokurov,$^{1}$
N.W. Evans$^{1}$
\\
$^{1}$Institute of Astronomy, University of Cambridge, Madingley Road, Cambridge, CB3 0HA, United Kingdom\\
}

\date{Accepted XXX. Received YYY; in original form ZZZ}

\pubyear{2020}

\begin{document}
\label{firstpage}
\pagerange{\pageref{firstpage}--\pageref{lastpage}}
\maketitle

\begin{abstract}
We assemble a catalogue of Magellanic Cloud red giants from Data Release 2 of the $Gaia$ mission and, utilising machine learning methods, obtain photometric metallicity estimates for them. In doing so, we are able to chemically map the entirety of the Magellanic System at once. Our high resolution maps reveal a plethora of substructure, with the Large Magellanic Cloud (LMC) bar and spiral arm being readily apparent. We uncover a curious spiral-like feature in the southern portion of the LMC disc, hosting relatively metal-rich giants and likely a by-product of historic encounter with the Small Magellanic Cloud (SMC). Modelling the LMC as an inclined thin disc, we find a shallow metallicity gradient of $-0.048 \pm 0.001$ dex/kpc out to $\sim 12^{\circ}$ from the centre of the dwarf. We see evidence that the Small Magellanic Cloud is disrupting, with its outer iso-density contours displaying the S-shape symptomatic of tidal stripping. On studying the proper motions of the SMC giants, we observe a population of them being violently dragged towards the larger Cloud. The perturbed stars predominately lie in front of the SMC, and we interpret that they exist as a tidal tail of the dwarf, trailing in its motion and undergoing severe disruption from the LMC. We find the metallicity structure in the Magellanic Bridge region to be complex, with evidence for a composite nature in this stellar population, consisting of both LMC and SMC debris. 
\end{abstract}




\section{Introduction}
The history of the Large and Small Magellanic Clouds is fraught with complex interactions and, being our nearest example of such a system, provides us with a vital laboratory for detailed study of interacting dwarf irregulars. The Large Magellanic Cloud (LMC) is generally well described by a planar, inclined disc yet still displays a host of deviations from this simple picture; it has long been observed to display one dominant spiral arm \citep[see e.g.][]{deVac_1955, deVac_1972} as well as an off-centred stellar bar \citep[e.g.,][]{Zhao_2000,Nikolaev_2004} and shell/ring like features \citep[e.g.][]{deVac_1955, Irwin_1991, Choi_2018_b}. Probing the planarity of the LMC using red clump (RC) stars, \citet{Olsen_2002} found an inner south-west region to display a prominent warp, curiously located in the portion of disc nearest to the Small Magellanic Cloud (SMC). More recently, \citet{Choi_2018} identified an outer warp in the LMC, departing from the LMC plane by up to $\sim 4$ kpc toward the SMC. Towards the outer regions of the LMC, a plethora of substructure is observed with \citet{Mackey_2016} first identifying a 10 kpc stellar arc located in the northern periphery of the LMC. The  optical imaging of the Clouds by \citet{Besla_2016} further revealed significant substructure in the northern potions of the LMC. The recent deep imaging of \citet{Mackey_2018}, in combination with that of the Dark Energy Survey (DES), revealed diffuse stellar substructures residing south of the LMC along with LMC disc truncation in the south and western regions \citep[see also][]{Clouds}. The state of the metal-poorer SMC is even more complex and disordered, exhibiting a roughly triaxial ellipsoidal shape with substantial depth along various lines of sight \citep[see e.g.][]{Gardiner_1991, Subramanian_2012,  Deb_2015, Scowcroft_2016, Jacyszyn-Dobrzeniecka_2017,  Muraveva_2018}. There is evidence of stellar debris tidally stripped by the LMC lying in its eastern regions \citep{Nidever_2013, Subramanian_2017}. Centrally, the SMC displays a less prominent bar as well as an eastern wing structure first observed by \cite{Shapley_1940}, departing from the northern end of the bar toward the LMC. Recent mapping of the SMC by \citet{Youssoufi_2019} has shown the morphology of the younger stellar populations to be highly irregular and largely limited to the SMC bar and eastern wing, tracing the perturbed gaseous reservoir of the SMC \citep{Stan_2004}, with recent star formation excepted to have occurred in these regions \citep[see e.g.][]{Irwin_1990, Youssoufi_2019}. The older stellar populations, on the other hand, display a much more homogeneous spatial distribution \citep[see e.g.][]{Zaritsky_2000, Haschke_2012, Jacyszyn-Dobrzeniecka_2017} and indeed are offset from the younger populations, as observed by \citet{Mackey_2018}. Such distinctions are suggestive of historic perturbations to the SMC's gas supply. Furthermore, \citet{Olsen_2011} discovered a population of giants residing in the LMC yet distinct in their kinematics with respect to the local field. This disparity was compounded by the fact that these stars were also significantly metal-poorer than would be expected for an LMC disc population, leading the authors to the conclusion that these stars in fact originated in the SMC, having been accreted onto the more massive LMC.

The general features outlined above clearly demonstrate that we cannot consider the Clouds independently. Their present day morphology is directly influenced by historic, mutual interactions with a striking example being the existence of the Magellanic Bridge (MB). The structure was first observed by \citet{Hindman_1963} as a continuous structure of neutral hydrogen linking the two Clouds. \citet{Irwin_1985} later discovered this gaseous bridge to be a site of recent star formation. They observed hundreds of blue main sequence stars to lie in the bridge region, coincident with the HI distribution. The young stellar bridge has also been observed more recently. For example, \citet{Dana2012} map the bridge with OB stars selected using a combination of UV, optical and IR photometry. \citet{Skowron_2014}, utilise OGLE's expansive coverage of the region, while \citet{Mackey_2017} trace young stellar associations, ages $\lesssim 30$ Myr, from the SMC wing to the outskirts of the LMC, suggesting that their spatial coincidence with the densest regions of HI gas between the Clouds points to these stars having formed in situ. In addition, there is the $\sim 200^{\circ}$ long Leading Arm and trailing Magellanic Stream (MS) features associated with the system; purely gaseous in nature, they are strongly indicative of tidal stripping of the Clouds. The simulations of \citet{Diaz_2012} and \citet{Besla_2012} demonstrate that a historic close encounter between the Clouds is able to reproduce the general large scale features of the Magellanic Stream and Leading Arm. To reproduce the general characteristics of the gaseous Magellanic Bridge, a more recent encounter seems to be required, with swathes of gas being stripped from the SMC into the MB region, hosting star formation, alongside stripped stellar debris. 

Evidence for a stripped intermediate/old stars in the MB has been seen numerous times \citep[see e.g.][]{Bagheri_2013, Nidever_2013, Skowron_2014, Carrera_2017}. The RR Lyrae stars selected from $Gaia$ DR1 by \citet{Vasily_2017} revealed the presence of tidal tails around both the LMC and SMC. They traced a bridge of stars connecting the two Clouds. Studying the 3D structure of this bridge, they found a dual nature; the bridge exhibits one component displaying a smooth distance gradient from the SMC to the LMC, with the other at the mean distance of the LMC, indicative that the old stellar bridge is composite in stellar make up. That is, a large portion of the old stellar bridge is the trailing tail of the SMC, tidally stripped to closer heliocentric distances and towards the LMC. The leading arm is largely compressed on-sky, and elongated along our line of sight. \citet{Vasily_2017} went on to posit that a significant portion of the bridge is composed of LMC material, tidally dragged into the MB region. The follow-up studies by \citet{Mackey_2018} and \citet{Clouds} have confirmed the presence of a messy mix of extra-tidal sub-structure in this intra-Cloud region, sometimes referred to as "old Magellanic Bridge". Based on Hubble Space Telescope (HST) observations, \citet{Zivick_2018} showed this close encounter occurred of order 150 Myr ago, with an impact parameter of less than 10 kpc. The extent of the LMC disc is large, with \citet{Mackey_2016} observing it to continue to $\sim 18.5$ kpc in the DES imaging of the Cloud. Thus it seems extremely likely that the LMC and SMC have undergone a recent direct collision. The Clouds appear to be on their first infall into the Milky Way \citep{Besla_2007}, consistent with the notion that it is LMC-SMC interactions that are the main driver of the substructure we observe. Indeed, the deep photometry from the Survey of the Magellanic Stellar History (SMASH) analysed by \citet{Ruiz_Lara_2020} found evidence for the long term stability of the LMC's spiral arm, dating its origin to more than 2 Gyr ago, and suggesting its independence from the MW interaction. 
\begin{figure}
\centering
\includegraphics[width=\columnwidth]{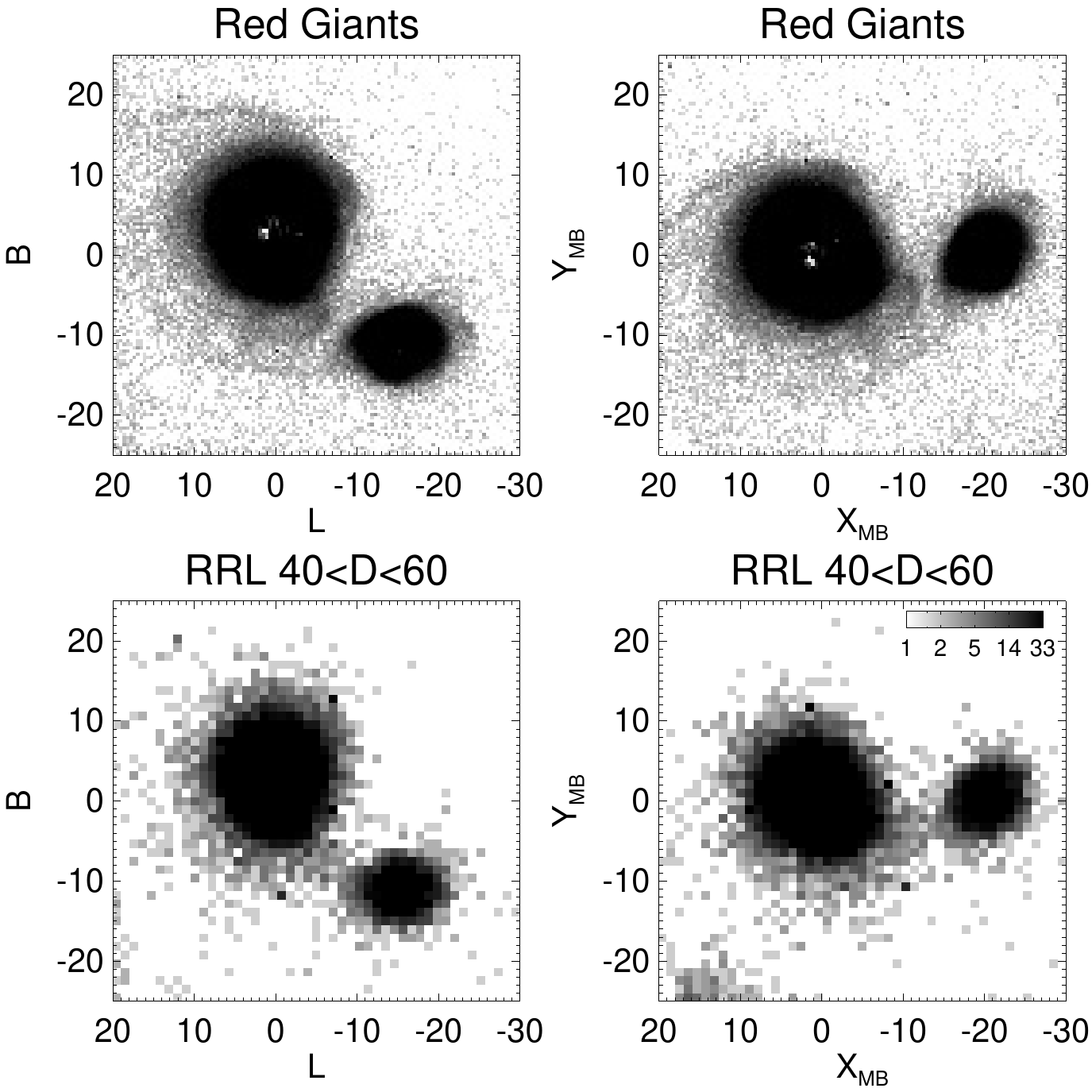}
    \caption[width=\columnwidth]{The top two panels show the red giant sample of \citet{Clouds} shown in both Magellanic Stream and Magellanic Bridge coordinates. A myriad of diffuse stellar substructure can be seen in the outer regions of the system. Most prominent is the northern spiral like feature that was first observed by \citet{Mackey_2016}, alongside a host of complex thin streams in the southern portions of the outer LMC. The bottom two panels display $Gaia$ DR2 RR Lyrae, whose selection we describe in the text. Note that the RR Lyrae distribution follows closely that of the giants, including the sharp cutoff in the disc density on the side of the LMC facing the SMC. With both stellar tracers, the old stellar bridge \citep{Vasily_2017} is evident as the connecting feature of the two Clouds.}
\label{fig:giant_rr}
\end{figure}

In this work, we focus on producing high resolution metallicity maps of the Magellanic Clouds. These provide a vantage point, from which we can observe the complex structural properties of the system. Generally speaking, the LMC is thought to have a shallow negative metallicity gradient through its disc, hosting a metal-rich bar centrally. \citet{Cioni_2009} utilised the ratio between C and M type asymptotic giant branch stars as an indicator for mean metallicity. Indeed, they detect a metallicity gradient of $-0.047 \pm 0.003$ dex/kpc in their sample, out to a radius of $\sim 8$ kpc from the centre of the dwarf. This is consistent with MC red giant analysis of \citet{Choudhury_2016}, whose spectroscopically calibrated photometric $\feh$ estimates revealed a gradient of $-0.049 \pm 0.002$ dex/kpc through the LMC disc out to $\sim 4$ kpc. They further find evidence for differing metallicity gradients in the northern and southern portions of the LMC disc, in comparison to the eastern and western regions, with the latter being relatively metal poorer. With regards to the SMC, the spectroscopic measurements \citet{Carrera_2008} indicated a negative metallicity gradient is also present in the smaller Cloud, with the recent analysis of \citet{Choudhury_2020} quoting a gradient value of $-0.031 \pm 0.005$ dex/kpc, derived from their photometric $\feh$ estimates for SMC red giants. Interestingly, they also saw evidence for asymmetries in the SMC metallicity profile with it appearing flatter towards the east, in the direction of the LMC, than in the west. They interpret this to be a potential indicator for stellar mixing, acting to flatten the gradient, likely stemming from LMC-SMC interactions. We will assemble a large collection of Magellanic red giants and, on assigning them photometric metallicities, map the Magellanic system in its entirety.

\section{Data}
\label{sec:data}
\begin{figure}
\centering
	\includegraphics[width=\columnwidth]{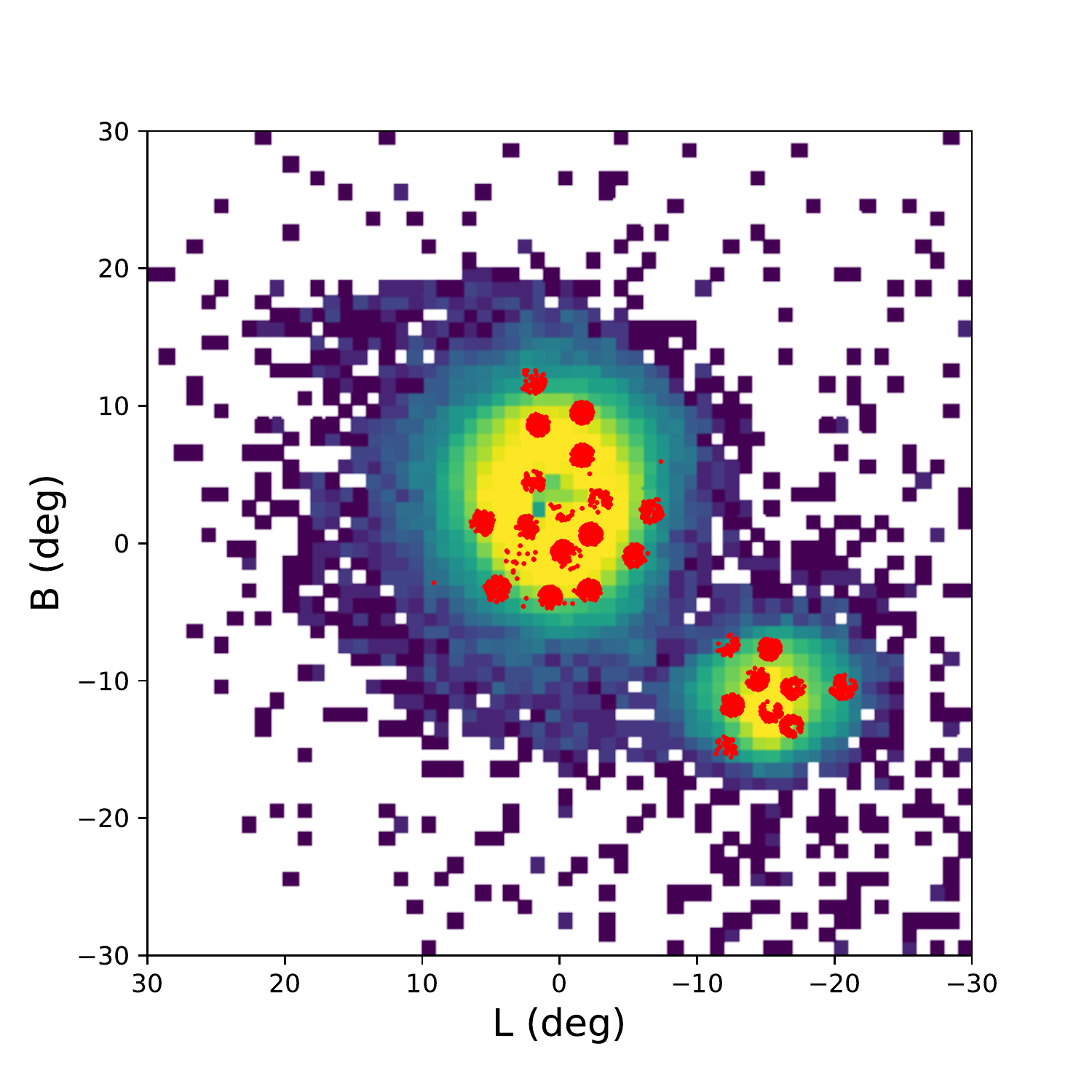}
    \caption[width=\columnwidth]{Spatial distribution of our giants in Magellanic Stream coordinates. Stars with an APOGEE spectroscopic metallicity measurement are shown in red and provide good coverage of both satellites. These stars will constitute our regression training set to predict the metallicities of all the Magellanic giants in our sample.}
\label{fig:training_LB}
\end{figure}
We make use of the Magellanic red giant (RG) catalogue compiled by \citet{Clouds} from Data Release 2 \citep{GDR2_2018} of the $Gaia$ mission \citep{Gaia_mission}, which holds 1\,604\,018 stars within $30^{\circ}$ of the Clouds. In the top row of Fig.~\ref{fig:giant_rr}, we show the stellar density of these giants in two coordinate systems: the Magellanic Stream system (L,B) described by \citet{Nidever_2008} and the Magellanic Bridge system as in \citet{Vasily_2017}, both of which will be utilised in this work. A plethora of substructure is seen in the giants, with the Clouds hosting multiple outer spiral-like arms such as the Northern structure identified by \citet{Mackey_2016}. A complex ensemble of features present in the southern portion of the LMC; the claw like features, associated with "Substructure 1" and "Substructure 2" of \citet{Mackey_2018}, appear to wrap clockwise around the lower LMC disc. As noted in \citet{Clouds}, one of the most striking features is the thin stellar stream that appears connected to the SMC, arcing $\sim 90^{\circ}$ clockwise around the outer LMC. The portion of the LMC disc nearest the SMC is truncated, with the morphology of the smaller dwarf appearing distorted, apparently stretching towards the LMC. The outskirts of the two galaxies are littered with stellar debris, and indeed a population of giants is observed to inhabit regions between the Clouds. In the lower two panels of the figure we show, for comparison, the distribution of RR Lyrae (RRL) in the two coordinate systems, combining both the $Gaia$ SOS \citep[Specific Object Study][]{Gaia_RRL_SOS} catalogue with stars classified as RRL in the general variability table $\texttt{vari\_classifier\_result}$ \citep{Gaia_vari_2018}. Requiring these RRL to have $\texttt{phot\_bp\_rp\_excess\_factor} < 3$, we correct for extinction and assign heliocentric distances as in \citet{Iorio_2019}.  Selecting those RRL whose distances are commensurate with the Clouds, we see in Fig.~\ref{fig:giant_rr} a relatively clean selection can be made. The peripheries of the dwarfs are again scattered with diffuse structures, the LMC disc is truncated towards the SMC and most strikingly of all is the old stellar bridge spanning in the inter-Cloud region. Interestingly, the RR Lyrae distribution follows closely that of the giants, implying that in the LMC, many of these old pulsating stars represent the disc population. This is perhaps best reflected by the sharp disc truncation on the side nearest to the SMC. Both red giants and RR Lyrae show this dramatic near-linear cutoff in the disc density. The evacuated western portion of the LMC's disc emphasises strikingly the old bridge connection between the two Clouds.

\begin{figure*}
\centering
	\includegraphics[width=\textwidth]{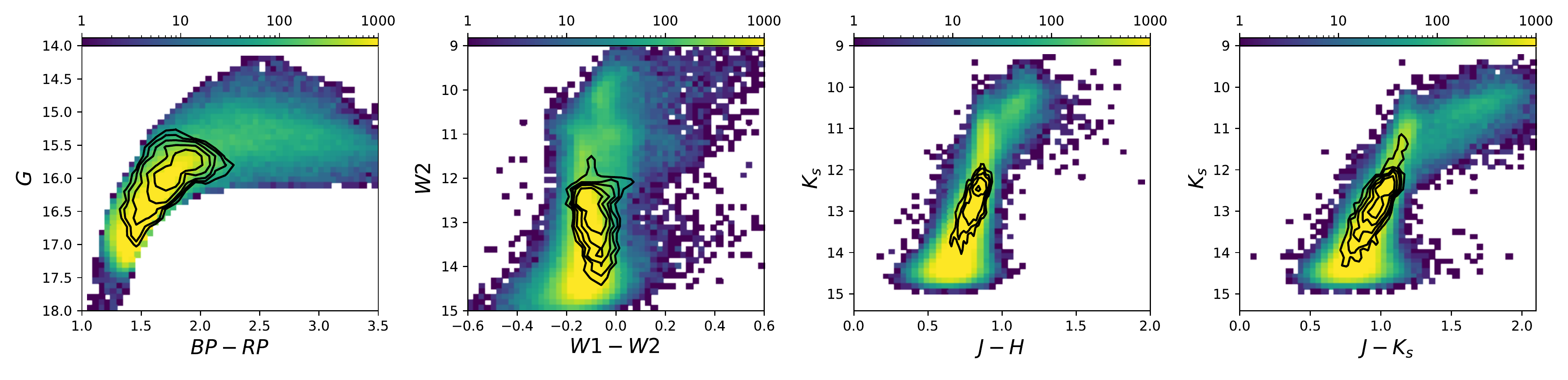}
    \caption[width=\textwidth]{CMD diagrams of the full $Gaia$+2MASS+WISE giants for which we will predict metallicities. We see our giant sample encompasses the upper RGB out to the AGB stellar phase. The cleanliness of the $Gaia$ CMD is a direct consequence of the methods employed by \citet{Clouds} in their giant selection. In the other photometric systems, the CMDs look relatively clean and we overlay black contours spanning the $5^{\textup{th}} - 50^{\textup{th}}$ percentile levels of the APOGEE training sample, with logarithmic spacing. It can be seen that APOGEE generally obtains metallicity estimates for the brighter red giants, with the sample being particularly deficient at blue WISE colours. How this pertains to our regression analysis is discussed later in this work.}
\label{fig:CMD}
\end{figure*}
\begin{figure*}
\centering
	\includegraphics[width=\textwidth]{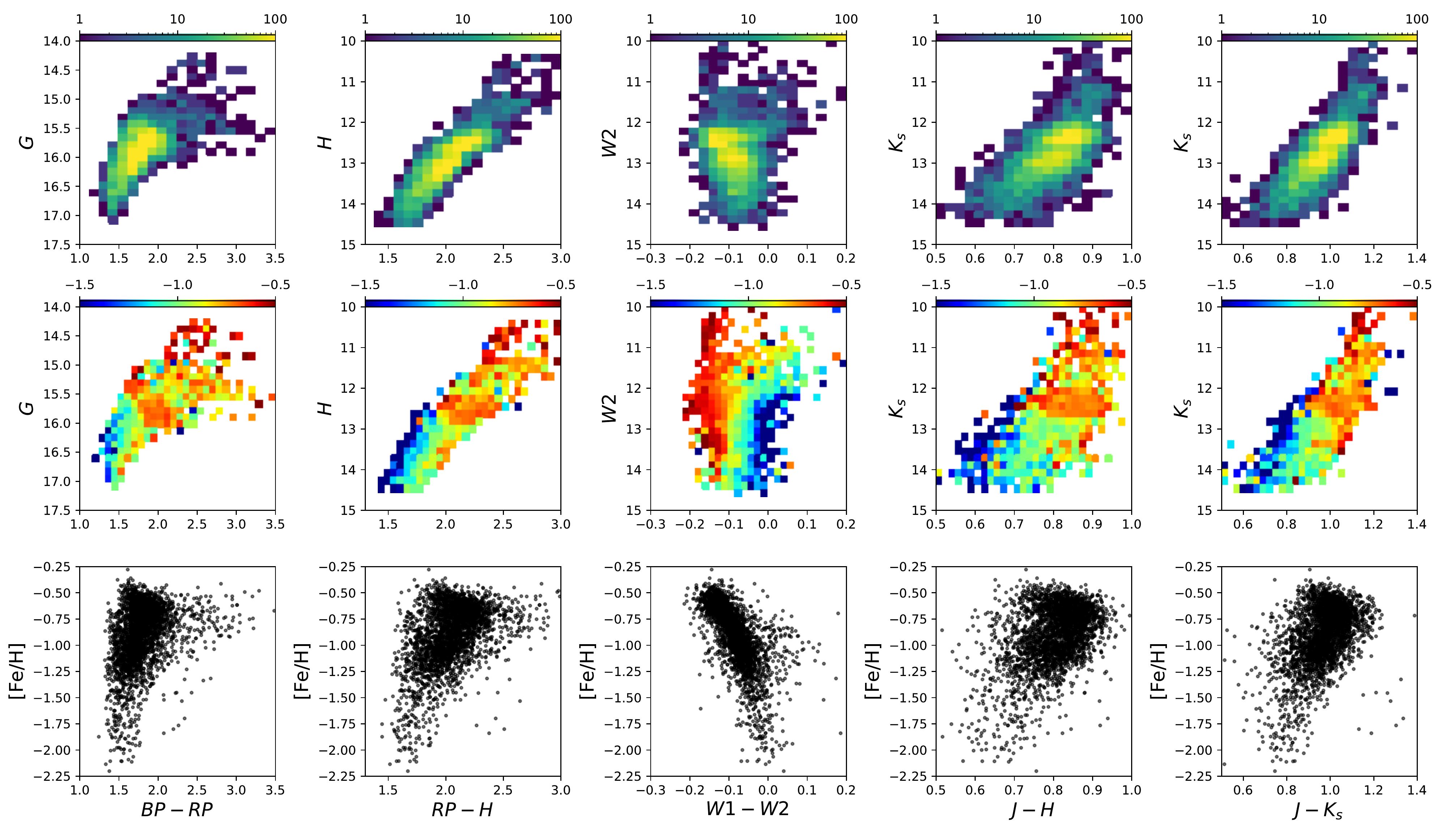}
    \caption[width=\textwidth]{\textit{Top row}: CMD of our APOGEE training set across the $Gaia$, 2MASS and WISE photometric systems utilised in this work along with the addition of the $Gaia$+2MASS combination chosen as a feature in our regression analysis. The bulk of stars lie in the upper RGB with few AGB stars present. \textit{Middle row}: The colour scheme here shows the mean metallicity in each pixel. Gradients are clear across all colours, especially so in WISE owing to its near vertical RGB. \textit{Bottom row}: Scatter plots between each of our four chosen features and APOGEE metallicity are shown. Positive correlations are evident with the WISE band showing the tightest relation.
    }
\label{fig:cmds}
\end{figure*}

 We cross-match the Magellanic giant sample the 2MASS \citep{2MASS_2006} and WISE \citep{WISE_2010} surveys to build a sample with broad photometric coverage. We only select stars with the 2MASS quality flags $\texttt{ph\_qual} = \texttt{\textup{AAA}}$, $\texttt{cc\_flg} = 000$, $\texttt{gal\_contam} = 0$ and the WISE quality flags $\texttt{ext\_flg} \leq 1$, $\texttt{ph\_qual} = \texttt{\textup{AAA}}$ to remove potential artefacts and sources with poor photometric measurements. Initially, we correct for extinction using the dust maps of \citet{Schlegel_1998} and the 2MASS and WISE extinction coefficients of \citet{Yuan_2018}. For the $Gaia$ photometry, we follow the de-reddening procedure of \citet{Gaia_2018}, using the first two terms in their equation (1) to do so. The recent red clump calibration of \citet{Skowron_2020} has provided reddening maps for the LMC and SMC with good resolution in the central parts of the Clouds, regions where those of \citet{Schlegel_1998} suffer from high levels of dust. Consequently, we utilise these recent maps for the innermost regions of the Clouds \footnote{We also applied this procedure to the RRL in Fig.~\ref{fig:giant_rr}}. We then require the giants to obey the relation of $\texttt{phot\_bp\_rp\_excess\_factor}
< 1.3 + 0.06 \: (\texttt{BP\_RP})^{2}$ and reject any stars that now lie outside the CMD selection box of \citet{Clouds}, after correcting the extinctions in the inner regions, yielding a sample size of 226\,119 Magellanic giants. 

A subset of our sample are captured by the APOGEE-2 southern hemisphere observations \citep{Majewski_2016, Zasokwski_2017}, the MC targets of which provide a relatively unbiased sample of stars spanning a large metallicity range of $\feh = -0.2$ dex down to $\feh = -2.5$ dex (see \citep{Nidever_2020}). This provides us with metallicity values for 3\,077 giants. Fig.~\ref{fig:training_LB} shows the spatial distribution of our giant sample, with red markers indicating those stars with APOGEE measurements. Our full sample spans the entirety of the Magellanic region with the APOGEE stars also exhibiting good spatial coverage of the Clouds in both the radial and azimuthal sense. We will utilise this sub-sample as a training set to build a regression model that can accurately predict the metallicities for our full giant sample. In Fig.~\ref{fig:CMD}, we show the colour-magnitude diagrams (CMDs) of our giants with the extent of the training APOGEE data indicated by the black contours. The WISE and 2MASS CMDs are relatively clean with clearly discernible red giant branches (RGB) and asymptotic giant branches (AGB) seen; the marked drop in stellar density at approximately $K_{s} < 12$ and $W1 < 12$ marks the transition to the AGB stellar populations of which our sample encompasses both the Oxygen rich (O-rich) and Carbon rich (C-rich) components of this phase. The stars for which APOGEE measurements exist are largely confined to bright RGs, with a noticeable lack of stars bluer than $\sim$ -0.2 in the WISE colour $W1-W2$ in comparison to the full sample. The effects that such colour offsets have on our ability to accurately predict metallicities is discussed later in this paper. APOGEE DR16 has observed $\sim 23\,000$ stars in the direction of the MCs, largely comprised of RGB, AGB and foreground dwarf stars \citep{Zasokwski_2017}. Recently, \citet{Nidever_2020} selected Magellanic giants from APOGEE by utilising optical photometry to remove foreground dwarfs, and devised a 2MASS photometric selection to isolate likely Magellanic giants. The authors acted to ensure there was minimal bias against selecting metal poor giants by employing a wide range in $(J-K_{s})$ selection. We probed our APOGEE sample in relation to that of \citet{Nidever_2020} by selecting all APOGEE+$Gaia$+2MASS+WISE stars within $40^{\circ}$ of the LMC as a comparison sample. We then applied our photometric cleaning cuts outlined earlier in this section, as well as requiring parallax values to be less than 0.2 and employed a proper motion cut similar to that in \citet{Vasily_2017}. We did not implement the CMD selection of \citet{Vasily_2017} but rather selected stars in the $(J-K_{s})$ vs. $H$ MC red giant selection box of \citet{Nidever_2020}. We then assessed the level of bias that our $Gaia$ CMD giant selection incurs against giant metallicity by assessing our sample completeness with respect to this comparison sample. In doing so, we estimate our sample completeness to be $\sim 80-90 \%$ across our full metallicity range and conclude that no significant bias against any particular metallicity is present in the data we will use to build our regression model.
\section{Regression Analysis}
\begin{figure}
\centering
	\includegraphics[width=\columnwidth]{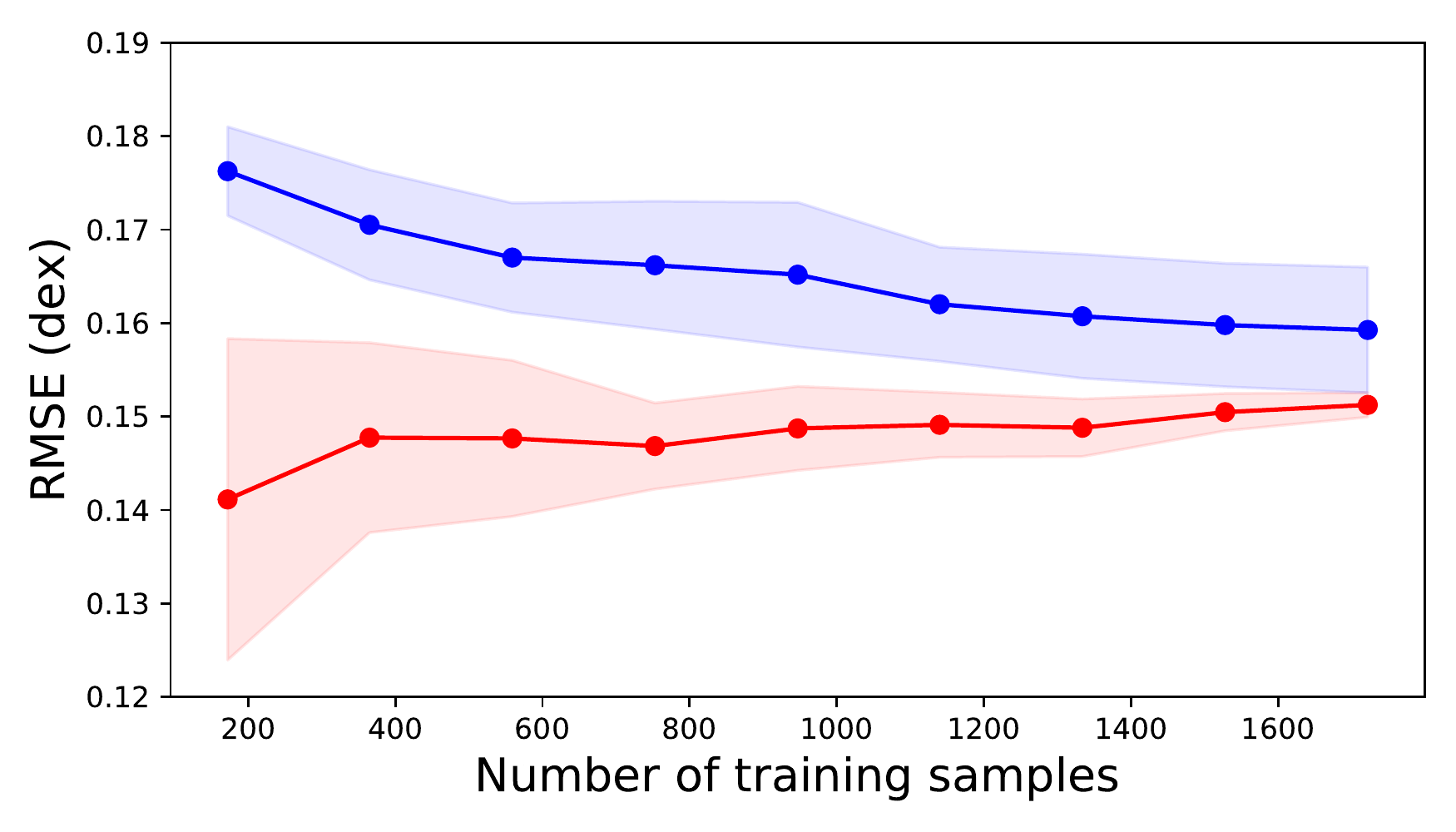}
    \caption[width=\columnwidth]{Learning curves for an SVR trained using a feature vector of $ \left [ \textup{BP-RP, W1-W2, J-H, J-K} \right ]$. A cross validation set is separated out and the algorithm is trained on incrementally increasing training set size. The root mean squared error (RMSE) score is then computed at each stage for both the training and validation set. Both curves tend toward each other smoothly to a sufficient accuracy of $\sim 0.17$ dex. The good convergence and low final accuracy indicates the learning scenario has sufficiently low bias and variance.}
\label{fig:LC}
\end{figure}
Through a process of experimentation, we choose a feature vector of $\textbf{f} = \left [BP-RP, RP-H, W1-W2, J-H, J-K_{s} \right ]$ to utilise in predicting photometric metallicities. The top row of Fig.~\ref{fig:cmds} shows the distribution of the APOGEE training set across five CMDs. We asses the photometric correlation with metallicity in the middle row of the figure, where across the feature vector we see clear metallicity gradients in the CMDs, especially so in the WISE photometry owing to its near vertical RGB. Computing the pair-wise correlations between training set features and metallicities yields the correlation vector $\left [ \textup{0.33, 0.45, -0.65, 0.46, 0.47} \right ]$, where elements are ordered in correspondence with feature vector $\textbf{f}$. These correlations are shown explicitly in the bottom row, where we see the strong correlation of the WISE colour with metallicity. The WISE $W2$ band covers the CO molecular absorption feature that is strongly dependent on stellar metallicity. Metal-rich giants are bright in the $W1$ band only whereas metal-poor giants are bright in both $W1$ and $W2$. Consequently, stars that are blue in $W1-W2$ are in fact metal richer, as seen in Fig.~\ref{fig:cmds} \citep[see e.g.][]{Schlaufmen_2014, Koposov_2015, Casey_2018}. The $Gaia$ features $BP-RP$ and $RP-H$ are included as both bear strong correlations with effective temperature, which itself affects the photometric colour of stars. We first set aside $30\%$ of our $Gaia$+APOGEE+2MASS+WISE sample as a test set, unseen by the learning algorithm throughout the training process and used for a final evaluation of the model's performance. Utilising the Support Vector Regression (SVR) implementation of $\texttt{scikit-learn}$ \citep{scikit-learn}, we implement a standard radial basis function (RBF) kernel of the form:
\begin{equation}
    K(\textbf{x}_{1}, \textbf{x}_{2}) = \textup{exp}\left ( -\gamma \left |  
    \textbf{x}_{1} - \textbf{x}_{2}
    \right |^{2} \right )
\end{equation}
and optimize the algorithm's parameters, namely $\gamma$ and regularisation parameter $C$, through a K-fold cross validation grid search, with 10 splits, accepting parameter values that minimise the root mean squared error (RMSE). We choose to use a SVR algorithm as it has the ability to model complex, non-linear relations between the features with relatively few parameters to tune.  The performance of the regressor is first assessed through the learning curve shown in Fig.~\ref{fig:LC} where the learning algorithm is trained on an incrementally increasing number of training points. A validation sample is set aside beforehand, from which we can evaluate the performance of the model with respect to our chosen metric (RMSE) at each incremental step. The general trend of the figure shows both curves tending towards a final RMSE of $\sim -0.16$ dex, an acceptably low value that validates the algorithm to be sufficiently unbiased. The convergence of the two curves also indicates relatively low variance - a regressor suffering from over-fitting would yield curves that tend toward convergence but still remain offset by a significant value even when maximal training data is available. 
%
%
\begin{figure*}
\centering
	\includegraphics[width=\textwidth]{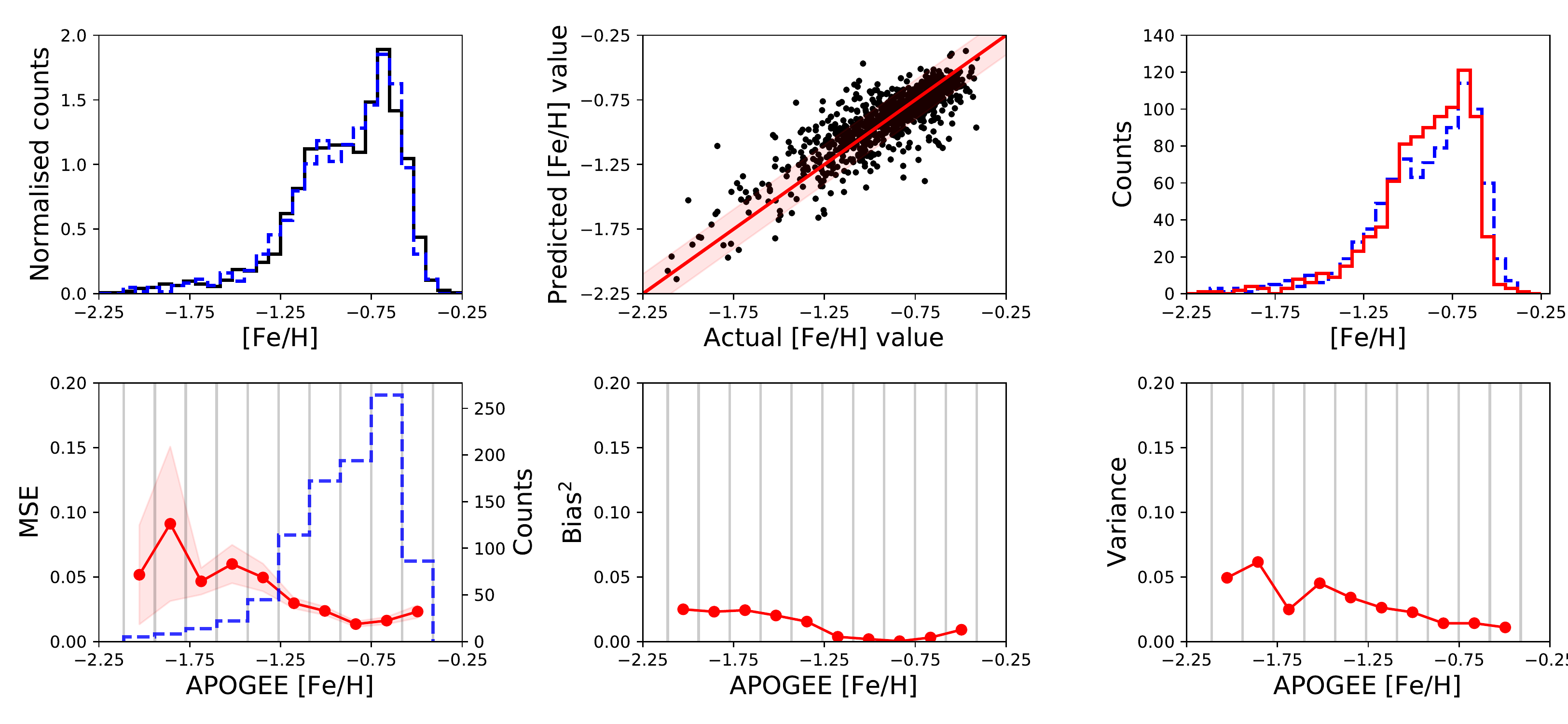}
    \caption[width=\textwidth]{\textit{Upper left panel}: Training data metallicity distribution in black solid line which is adequately representative of the test data distribution shown as blue hashed line. \textit{Upper middle panel}: Scatter plot of the predicted and APOGEE measured metallicities for the stars in our test set. The solid red line indicates a one to one correspondence and the red shaded region bounds the RMSE of 0.15 dex recovered from the test set analysis. The majority of the scatter is present at the lowest of metallicities where the stellar density is extremely low. \textit{Upper right panel}: The blue hashed line indicates the test set's true metallicity distribution now overlaid with the predictions in solid red. Generally, the overall distribution is reproduced well. \textit{Lower left}: The red line shows the estimated MSE binned by APOGEE metallicity values of our test data. The filled shaded region bounding this line shows the dispersion of this value weighted by the Poisson noise in each bin. The blue histogram shows the metallicity distribution of the test set in question with the faint grey vertical lines indicating the bin edges. At metallicities less than $\sim -1.2$ dex, the corresponding RMSE is 0.23 dex. \textit{Lower middle}: The squared bias in each metallicity bin is seen to be relatively low across the full range with the greatest contribution being at the metal-poor end. \textit{Lower right}: The variance of the predictions in each bin is shown. The largest contribution to total error is again seen at the metal-poor end owing to the dearth of training data in this region.}
\label{fig:test_preds}
\end{figure*}
\subsection{Regression Performance}
The upper middle panel of Fig.~\ref{fig:test_preds} shows good one to one agreement between the APOGEE spectroscopic $\left [ \mathrm{Fe}/\mathrm{H} \right ]$ values and those predicted by our trained regressor. The upper right panel of the figure shows we are able to reasonably reproduce the metallicity distribution of the test data. Whilst it appears that the regressor struggles at the LMC-SMC boundary of $\sim$ -0.8 dex from this figure, we stress that our aim is to accurately predict the metallicities of individual stars, not to reproduce the overall distribution. The lower panels of the figure shows the mean squared error (MSE), squared bias and variance as a function of APOGEE metallicity for the predictions on our test set. The model suffers from relatively little bias across the full metallicity range, with most contribution occurring at $\feh < -1.25$ dex. Generally, the main contribution to the total error stems from variance in the predictions. This is particularly the case at the low metallicity end where data are sparse, leading the model to over-fit in this region. The predictions are biased high by $\sim 0.15$ dex for metallicities less than -1.5 dex and are biased slightly low by $\sim 0.1$ dex at the most metal-rich end. For metallicities less than -1.2 dex, the RMSE is approximately 0.2 dex and the RMSE across all metallicity values is 0.15 dex. We list our prediction errors in four metallicity bins in Table.~\ref{table:errors}. we note that these are purely nominal error estimates and likely underestimated in some cases. However, this work will focus on the large scale metallicity patterns across the Clouds in which uncertainty in the predictions of individual stellar metallicities will have a small influence. 
\begin{table}
\centering
\begin{tabular}{ |c|c| } 
\hline
 $\feh$ bin & $\sigma_{\feh}$ (dex)  \\
 \hline 
 $\feh \leq$ -1.5 & 0.25 \vspace{1mm} \\ \vspace{1mm}
 $-1.5 < \feh \leq -1$ & 0.18 \vspace{1mm} \\ \vspace{1mm}
 $-1 < \feh \leq -0.5$ & 0.13 \vspace{1mm} \\ \vspace{1mm}
 $\feh > -0.5$ & 0.21 \vspace{1mm} \\ 
\end{tabular}
\caption{Table lists the RMSE, computed from our test set predictions, in four metallicity bins. It is in the tails of the metallicity distribution that our predictions suffer greatest from total error, as demonstrated in Fig.~\ref{fig:test_preds}. }
\label{table:errors}
\end{table}
\subsection{Regression Predictions}
We use our trained SVR to predict metallicities of our full $Gaia$+2MASS+WISE giants, whose spatial distribution is shown in Fig.~\ref{fig:training_LB}. We consider the fact that our full giant sample encompasses a much wider colour range than that of our APOGEE subset, particularly so for $W1-W2$ as seen in comparison of Fig.~\ref{fig:CMD} with Fig.~\ref{fig:cmds}. To account for this, we adopt a nearest neighbour approach, computing the mean Euclidean distance in feature space from each of our giants to those in the APOGEE sub-set using 5 nearest neighbours. We can then choose an appropriate upper limit cut on this mean feature distance, $\left \langle \mathrm{D_{NN}} \right \rangle $, to reject stars whose photometry is not well represented in the training data. We also impose cuts in the 2MASS colours for which we require $J-H < 1$ and $J-K < 1.25$.
\begin{figure*}
\centering%
	\includegraphics[width=\textwidth]{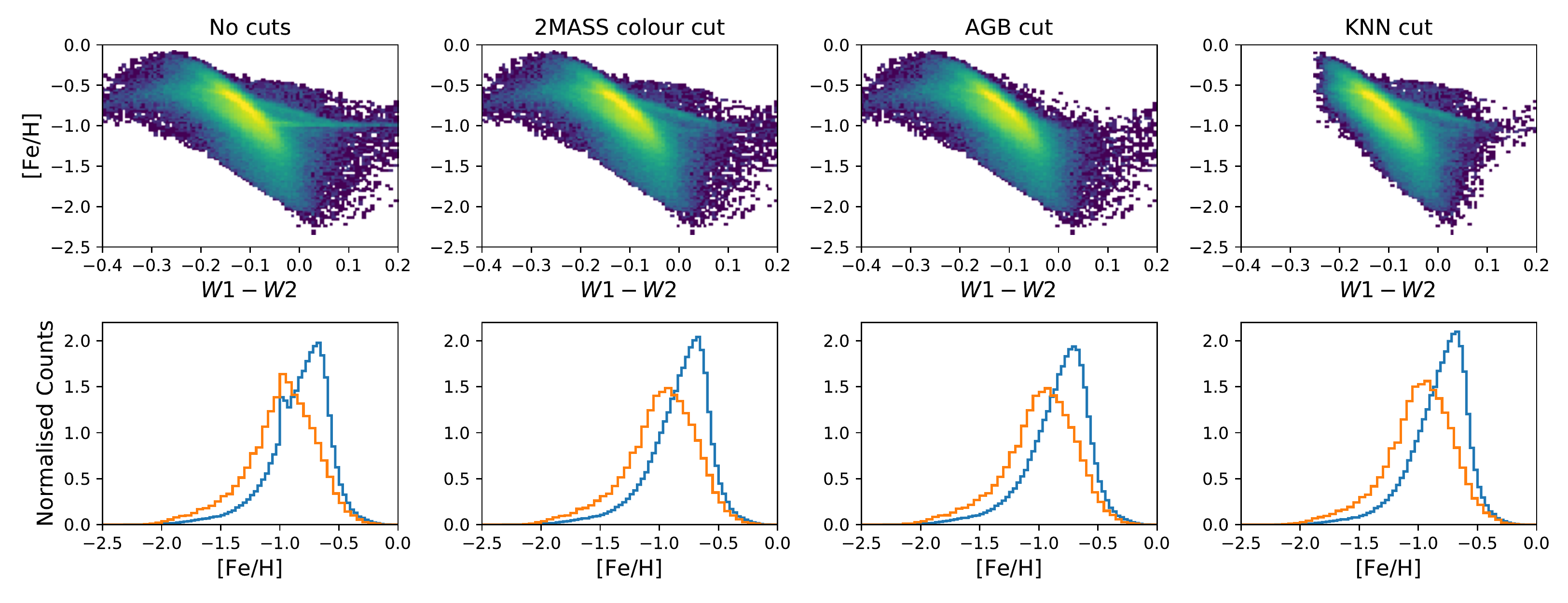}
    \caption[width=\textwidth]{Predicted metallicities for our $Gaia$+2MASS+WISE giants as a function of WISE colour $W1-W2$ are shown in the top row. The bottom row shows histograms of the predictions for LMC (SMC) in blue (orange) selecting stars that fall within a 12 (8) degree aperture of the respective Cloud. In the leftmost panels, there is a population of stars for which our predictions are spurious, causing a pile up at $\sim -1 $ dex. These stars are the reddest stars in our sample and can be removed through application of the 2MASS colour cuts described in the text. The AGB stars form a bifurcation in the metallicity-colour sequence owing to the unusual properties of the WISE isochrone tracks in which the AGB stars sharply turn red, crossing RGB tracks of lower metallicities. In turn, this causes a bi-modality in metallicity at fixed WISE colour. The final column shows the effect of requiring the mean feature space distance $\left \langle \mathrm{D_{NN}} \right \rangle $ to be less than 0.15. It largely removes the stars that are extremely blue, causing $W1-W2$ to be less than 0.15 and culling the bluest stars whose predictions spuriously converge to $\sim -0.8$ dex.}
\label{fig:cleaning_cuts}
\end{figure*}
In effect, this selection rejects the very red, likely C-rich, AGB stars from our sample. Such extremely red stars are poorly represented in the APOGEE training data and so the regression model will likely struggle to accurately interpolate values to this regime. We also further investigate the effect of removing the AGB branch by only considering stars with $K_{s} > 12$ (this selection is tuned to the LMC AGB as it is the dominant contributor to our sample). The effects of these individual selections are shown in Fig.~\ref{fig:cleaning_cuts}, which shows the predicted metallicity as a function $W1-W2$ colour for our $Gaia$+2MASS+WISE giants. The leftmost panel clearly shows spurious predictions accumulating at $\left [ \mathrm{Fe}/\mathrm{H} \right ] \sim -1$ dex, towards red values of $W1-W2$. On applying our 2MASS colour cuts we eliminate these entirely, as we cull the extremely red giants from our sample. The selection of stars fainter than $K_{s} > 12$ highlights the region occupied by likely AGB stars which consists of a spur branching out from the main colour-metallicity relation, an effect caused by the WISE AGB isochrone tracks turning red and crossing the RGB track at other fixed metallicities (see Fig. 1 of \citet{Koposov_2015}). The rightmost panels in the Figure show the effect of requiring the nearest neighbour mean feature space distance $\left \langle \mathrm{D_{NN}} \right \rangle $ to be less than 0.06, corresponding to a cull of stars within the largest $10^{th}$ percentile. It acts to mostly remove stars that are extremely blue in $W1-W2$ and whose predicted metallicities spuriously converge to $\sim -0.8$ dex at the blue end. Such stars are poorly represented in the APOGEE sample and so our model is again struggling to interpolate accurately in this regime. The bottom panels of the figure show our predicted metallicity distribution function (MDF) in each of the cases. Aside from the most spurious defections, these distributions behave well with both the LMC and SMC exhibiting appearing negatively skewed with long tails towards the metal-poor end. Hereafter, we will only consider giants in our analysis whose 2MASS colours obey they cuts $J-H < 1$ and $J-K < 1.25$, as it is this selection that eliminates the most serious artefacts in our predictions, yielding a sample of 218\,077 giants. We will note explicitly in the text when we apply further cleaning criteria to our giant sample. Considering stars within a $12^{\circ}$ aperture of the LMC, we obtain a median metallicity value of -0.78 dex, with the peak of the (skewed) distribution occurring at $\sim$ -0.67 dex. For stars falling within a $6^{\circ}$ aperture of the SMC, we obtain a median metallicity value of -0.96 dex, with the distribution peaking at $\sim$ -0.93 dex.

\section{Metallicity Maps}
\begin{figure*}
\centering
	\includegraphics[width=\textwidth]{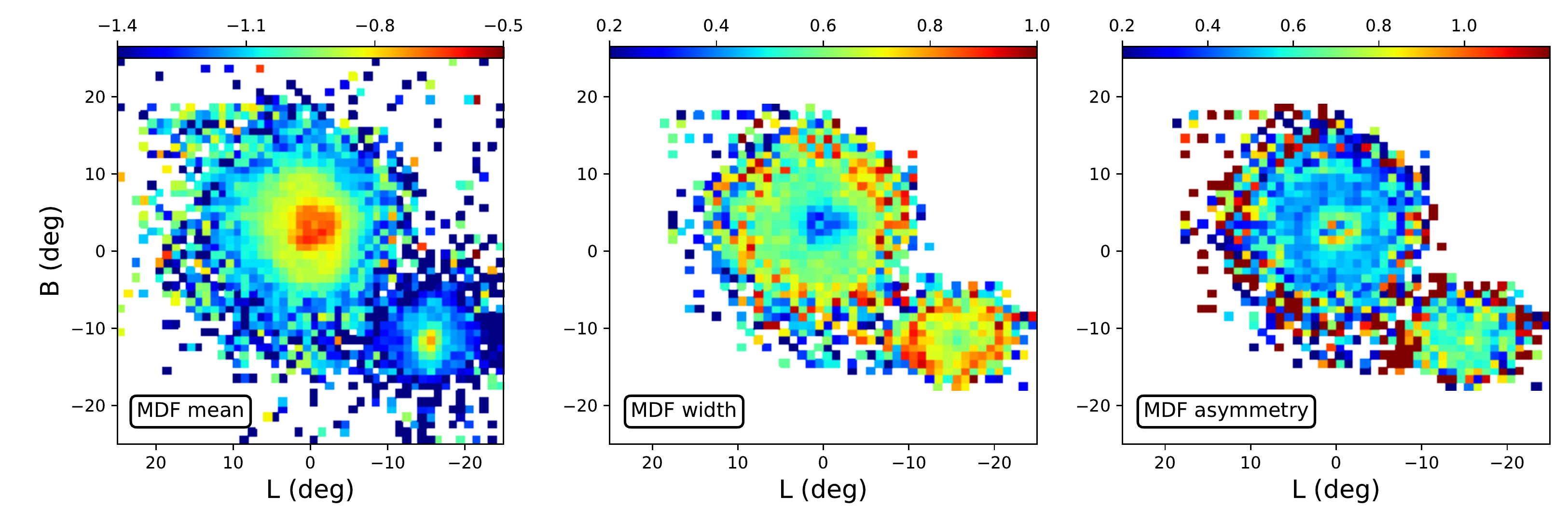}
    \caption[width=\textwidth]{\textit{Left}: Average metallicity maps of the Clouds coloured by mean metallicity per pixel. The central regions of the LMC are metal-rich, with the region occupied by the LMC bar and dominant spiral arm. Outside the bar region, diffuse metal-rich structures are seen with with arc-like morphology in both the southern and northern regions. A negative metallicity gradient is seen through the LMC disc with apparent asymmetry; the northern and southern portions of the dwarf appear to have relatively more metal-rich stars at larger radii, and hence a flatter gradient, as opposed to the east-west direction. The outermost regions of the LMC are littered with metal-poor stars. Our metallicity predictions show the SMC to be distinctly more metal-poor, with a slight gradient and central enhancement visible. \textit{Middle}: We show the difference between the $90^{\textup{th}}$ and $10^{\textup{th}}$ percentile of $\feh$ in each pixel, effectively a width measure of the MDF. The central regions of the LMC appear to have a relatively narrow MDF where the metal-rich bar and spiral arm dominates. Generally, the MDF of the SMC appears to be broader throughout the dwarf, in comparison to the LMC, likely owing to its large extent along the line of sight such that we are viewing a projection of multiple populations within the galaxy. \textit{Right}: We colour pixels by the ratio of the difference in the $95^{\textup{th}}$ and $50^{\textup{th}}$ to the difference in the $50^{\textup{th}}$ and $5^{\textup{th}}$ metallicity percentiles. A symmetric distribution would have a ratio of unity, with a left (right) skewed having values below (above) this. It can be seen that regions near the LMC bar have values close to unity, a consequence of metal-rich stars being relatively more prevalent in this region acting to reduce the left-skew of the MDF.  }
\label{fig:feh_map}
\end{figure*}
We show the mean predicted metallicity across the Clouds in the left panel of Fig.~\ref{fig:feh_map}, using the Magellanic Stream coordinate system of $(\textup{L}, \textup{B})$. The LMC bar is strikingly clear as a central metal-rich structure. Outside of the bar region, diffuse arcs and spiral arm-like structures are seen both in the northern and southern portions of the LMC disc. The LMC metallicity profile then decays into the outskirts, where it is littered with more metal-poor stars. The SMC is seen to be predominately metal-poor with only the most central region showing enhancement. The majority of substructure in the outskirts of the Clouds also appears to be relatively metal-poor, with the northern stream like substructure, identified first by \citet{Mackey_2016}, apparent in the map. The Clouds appear to be connected, largely with metal-poor stars, in two regions: that of the Magellanic Bridge consisting of stars likely stripped from the SMC and dragged toward the LMC as indicated by the proper motion analysis of \citet{Schmidt_2020}, and also in a region south of this at $\textup{L} \sim 8^{\circ}$, at the end of a narrow tail like substructure that wraps around the eastern edge of the LMC (see Fig.~\ref{fig:training_LB} and Fig. 2 of \citet{Clouds} also). Denoting the $i^{\textup{th}}$ percentile of the LMC/SMC MDF as $p_{i}$, we show in the middle panel the difference of $p_{90}-p_{10}$ which represents the width of the distribution. The central regions of the LMC display a consistently narrow MDF with that of the SMC generally quite broad, likely a projection effect of its extensive line of sight depth. In the right panel, we show the ratio of $(p_{95}-p_{50})$/($p_{50}-p_{5}$) which provides a sense of the direction of skew that the MDF possesses. 
The most central regions of the LMC show values close to unity, i.e. near symmetric, as metal rich stars dominate in this region and counteract the inherent left skew of the MDF (see Fig.~\ref{fig:cleaning_cuts}). The outskirts of the Clouds show a tendency for the MDF to tend towards symmetry, which may be an indication of relatively metal-richer stars originating in the inner disc having migrated outward. This is particularly the case for the eastern most edge of the SMC indicated by the region of black pixels at a MS longitude of $\sim -8^{\circ}$. Curiously, this region coincides (in projection) with the point at which the outer southern LMC spiral arm structure appears to join with the SMC (see Fig.~\ref{fig:giant_rr}). On closer inspection, we see there is a population of giants with $\feh > -1$ dex in this region, relatively metal-rich for outer SMC stars which generally take on $\feh < -1.3$ dex in our sample. It is plausible therefore that there exists a stripped LMC giant population in the region, which is the continuation of the outer spiral-like arm seen in Fig.~\ref{fig:giant_rr}.

Considering both Clouds individually, we show mean metallicity, stellar density and associated extinction maps in Fig.~\ref{fig:lmc_smc_maps}. Here, we show our giants in coordinates offset from the respective Clouds, adopting an LMC centre of $(\alpha_{0}, \delta_{0}) = (82.25^{\circ}, -69.50^{\circ})$ as determined by \citet{vdm_2001} and an SMC centre of $(12.60^{\circ}, -73.09^{\circ})$ from \citet{Rubele_2015}. Again the central bar is prominent and appears to display an extended metal-rich association just north of it; this extension is most likely the main spiral arm of the LMC. On slightly decreasing the dynamic range of the pixel colour, a plethora of diffuse metallicity features within the LMC is revealed in the figure, notably the strikingly spiral-like feature in the southern portion of the disc, reaching down to $\sim 6^{\circ}$ below the LMC centre. Its morphology is relatively smooth and coherent until $(\Delta \textup{L}, \Delta \textup{B}) \sim (2^{\circ},-5^{\circ})$, beyond which the metallicity structure becomes clumpy. This spiral-like feature, whilst faint, can also be seen in the corresponding stellar density map. The lack of correlation with any large scale extinction patterns, seen in the right most panel, supports the notion of this being a genuine feature of the LMC disc. When comparing our LMC metallicity map with that of \citet{Choudhury_2016}, the large scale features are generally consistent. In their analysis, they combined giants from the Optical Gravitational Lensing Experiment (OGLE-III) and Magellanic Cloud Photometric Survey (MCPS) out to a radius of $\sim 5^{\circ}$. Taking the slope of the RGB as proxy for average metallicity, calibrated against spectroscopic data, they estimated the metallicity of sub-regions through the Cloud. Both samples provide reasonable coverage of the central LMC with the OGLE-III (MCPS) footprint covering more of the east-west (north-south) regions. They observed the LMC bar to be the most metal-rich region and found evidence for a differing metallicity gradient through different regions of the disc; a shallower gradient was observed in the north-south regions in comparison to that in the east-west regions of the LMC. Both of these features are evident in Fig.~\ref{fig:lmc_smc_maps}, where the metal enhanced substructures in the top left panel are seen to reside largely in the north-south direction where they act to flatten any pre-existing gradient. With regards to the SMC, it is only the core of the dwarf that displays any coherent metallicity structure, consistent with the spatial density patterns observed by \cite{Youssoufi_2019} and the recent SMC metallicity maps of \citet{Choudhury_2020}. The outskirts of the SMC appear to be stretched and show an elliptical appearance, likely a result of tidal stripping of material through LMC interactions \citep[see also][]{Vasily_2017, Massana_2020}. The black hashed (dotted) lines in the figure denote selection bounds to sample giants lying along the projected major (minor) axis of the Magellanic bars for use in Section~\ref{sec:grads}. Owing to the complex three dimensional structure of the SMC, the exact orientation at which we are viewing it is highly uncertain, and so in this case, the bar selection is made simply to isolate the central most metal-rich feature. It is worthwhile to note that, whilst the stellar density of our data-set in the central regions of the Clouds is patchy, we can still gain insight into the structure of these regions through the lens of spatially averaged metallicity maps. 

\subsection{Metallicity Gradients}
\label{sec:grads}
\begin{figure*}
\centering
	\includegraphics[width=\textwidth]{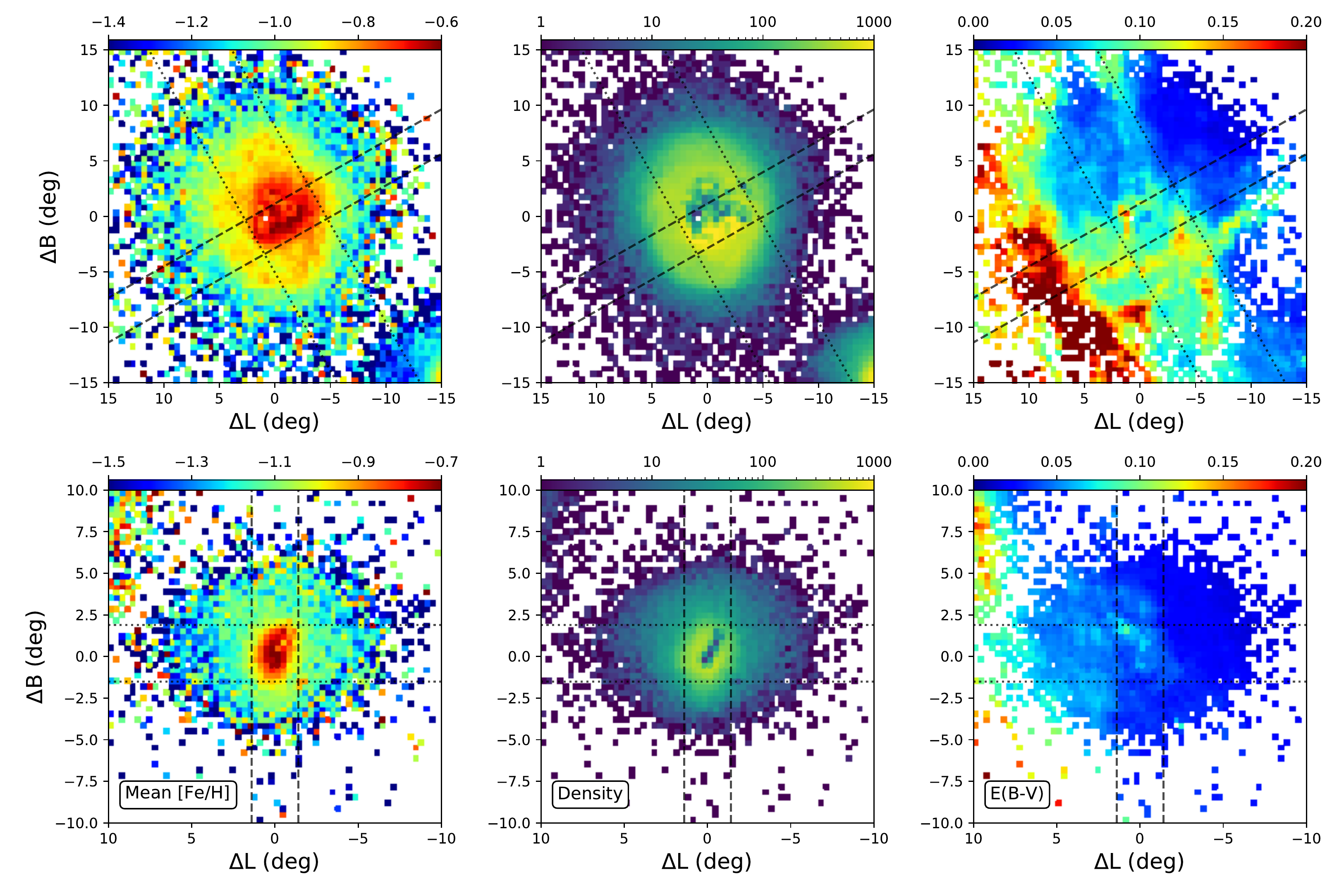}
    \caption[width=\textwidth]{Top (bottom) row of the Figure corresponds to the LMC (SMC). The left column shows mean metallicity per pixel. The same general features are seen as in Fig.~\ref{fig:feh_map} with the central regions of both Clouds being the most metal-rich and diffuse, metal enhanced structures seen in the outer LMC disc. The middle column shows the stellar density map of the Clouds, central regions of high extinction are most affected yielding gaps in the density profile. The SMC displays a stretched morphology with wings on the eastern and western side of the dwarf, a consequence of the tidal interactions with the LMC. The right columns shows the extinction map adopted in this work. Generally, the extinction is low aside from the eastern side of the LMC lying nearest the Galactic plane and various filamentary structures throughout it. Stars that fall between the dashed (dotted) lines are those shown in Fig.~\ref{fig:profiles} to trace the metallicity profile of the LMC and SMC along the projected major and minor axes of their respective bars.} 
\label{fig:lmc_smc_maps}
\end{figure*}
\begin{figure}
\centering
	\includegraphics[width=\columnwidth]{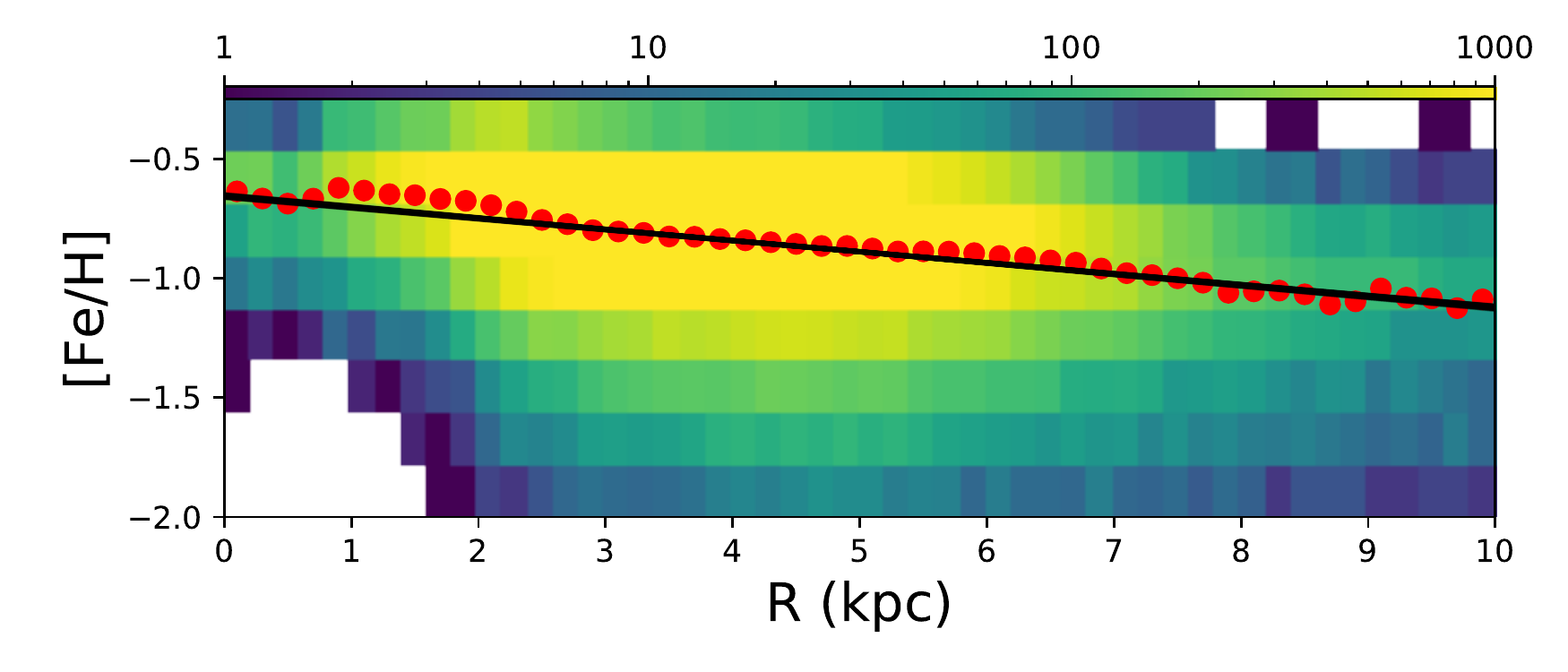}
    \caption[width=\columnwidth]{This shows the evolution of LMC giant predicted metallicity as a function of in-plane cylindrical radius. Pixels are coloured by stellar counts per pixel. The central regions are seen to be metal enhanced with respect to the outer regions of the LMC, where the metal-rich bar dominates. The red markers show the mean metallicity in bins of radius. A shallow negative metallicity gradient is evident.
    The solid black lines are samples drawn from our model. We only show giants with $\left \langle \mathrm{D_{NN}} \right \rangle $ < 0.06 in this figure to isolate stars that should have reasonable metallicity estimates given their colours. In this figure, we adopt the LMC viewing angles of \citet{Choi_2018}}
\label{fig:lmc_grad}
\end{figure}

We first attempt to quantify the presence of a metallicity gradient within the LMC by considering the Cloud as an inclined thin disc. The equations of \citet{vdm_2001} provide the transformations into a Cartesian system of an inclined thin plane, from observed on-sky positions, defined by its inclination angle $i$ and position angle $\theta$ measured anticlockwise from west \footnote{Note that the usual astronomical convention is to measure position angle from north and is related to $\theta$ by PA = $\theta$ - 90}. This allows us to assign each LMC giant an in-plane Galactocentric cylindrical radius $R$. We model the radial metallicity profile by the simple linear relationship:
\begin{equation}
\left [ \mathrm{Fe}/\mathrm{H} \right ]^{\textup{model}} = \Theta_{0}R(\alpha, \delta, i,\theta) + \Theta_{1}
\label{eq:model}
\end{equation}
where we wish to infer the gradient and intercept contained in the parameter vector $\boldsymbol{\Theta} = [\Theta_{0},\Theta_{1}]$. The in-plane radius is a function of the plane geometry and thus dependent on the choice of $(i,\theta)$ for the LMC. We account for this in our inference by marginalising over these parameters. We fix the centre of the LMC to be at $(\alpha_{0}, \delta_{0}) = (82.25^{\circ}, -69.50^{\circ})$ as in \citet{vdm_2001} and specify a distance of 49.9 kpc \citep{deGrijs_2014} to the LMC centre. Writing the total likelihood  as:
\begin{multline}
    p(\left [ \mathrm{Fe}/\mathrm{H} \right ] | \Theta, i, \theta,\alpha,\delta,\sigma_{ \mathrm{Fe}/\mathrm{H}}, V) = \\ \prod_{n}^{N} \frac{1}{\sqrt{2\pi s_{n}^{2}}} \; \textup{exp} (-\frac{(\left [ \mathrm{Fe}/\mathrm{H} \right ]^{\textup{model}}_{n}-\left [ \mathrm{Fe}/\mathrm{H} \right ]_{n})^{2}}{2s_{n}^{2}})
\end{multline}
where $\left [ \mathrm{Fe}/\mathrm{H} \right ]_{n}$ is our metallicity prediction for the $n^{\rm th}$ giant. The term $s_{n}^{2} = \sigma_{n}^{2} + V$ encompasses the prediction error $\sigma_{n}$ for each giant and some general intrinsic scatter in the model through the parameter $V$. The full posterior probability can then be written as:
\begin{multline}
    p(\Theta ,V | \alpha, \delta, \left [ \mathrm{Fe}/\mathrm{H} \right ], \sigma_{ \mathrm{Fe}/\mathrm{H}}) = \\
    \int \int p(\left [ \mathrm{Fe}/\mathrm{H} \right ] | \Theta, i, \theta,\alpha,\delta,\sigma_{ \mathrm{Fe}/\mathrm{H}}, V)\,p(\Theta, V)\,p(i,\theta)\,di\, d\theta
\end{multline}
from which we can draw samples in an MCMC fashion, utilising the sampler \texttt{emcee} of \citet{emcee}, for parameters $\Theta$ and $V$. 
\begin{figure*}
\centering
	\includegraphics[width=\textwidth]{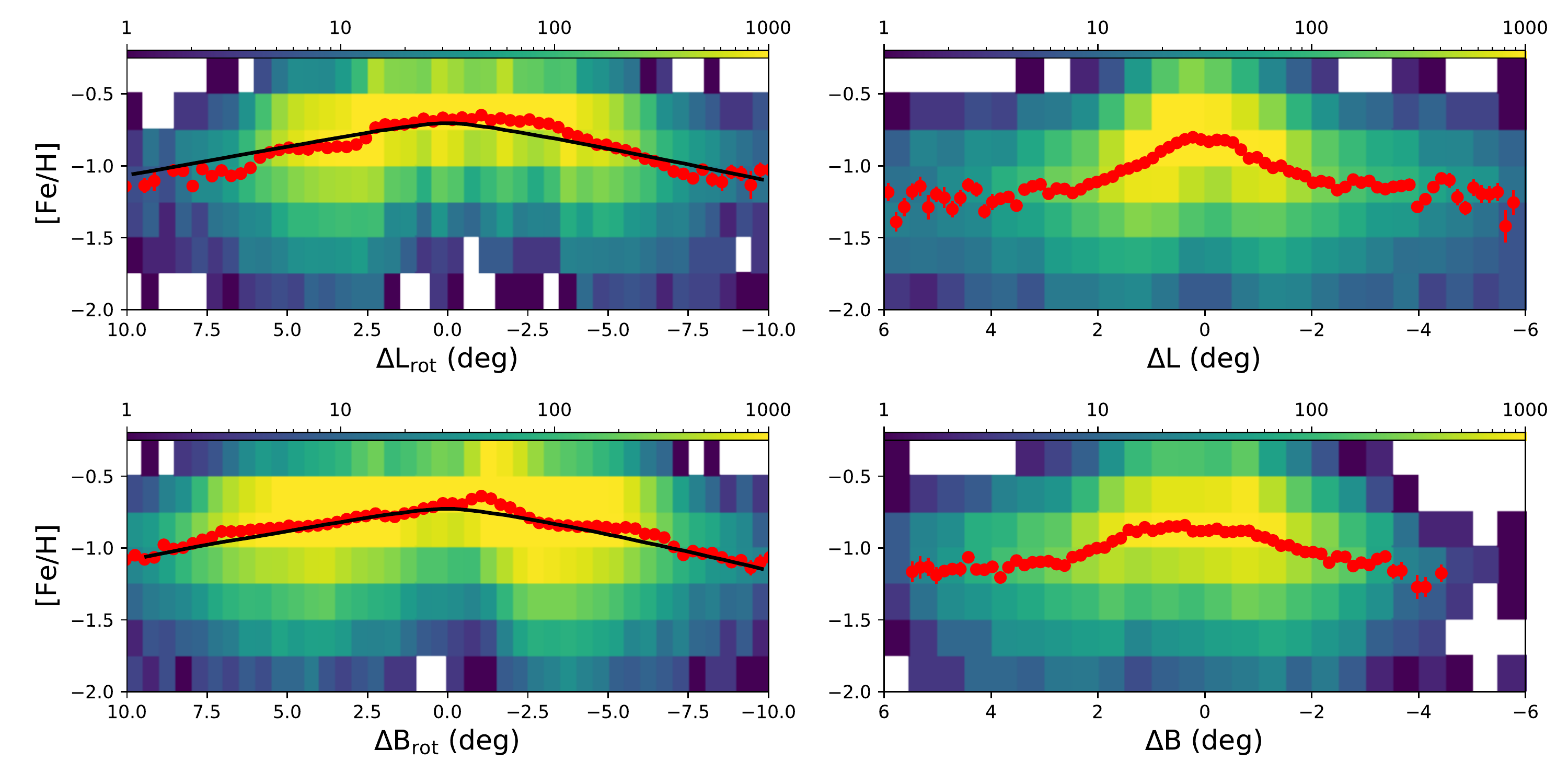}
    \caption[width=\textwidth]{Stellar density maps tracing the metallicity profile along directions aligned with the projected axes of the bars. The left two panels corresponds to the LMC and the right panels to the SMC. The top (bottom) row shows the profile of stars that fall within the dashed (dotted) line in Fig.~\ref{fig:lmc_smc_maps}. In the case of the LMC, we have rotated the Magellanic Stream coordinate system clockwise by $30^{\circ}$ so as to approximately align with the major axis of the bar. We only consider stars with $\left \langle \mathrm{D_{NN}} \right \rangle < 0.06$ in this Figure. Red markers correspond mean metallicity binned over angular offset relative to each Cloud's centre with error bars representing standard errors in each bin. Both the LMC and SMC show a flattening in the metallicity profile centrally in the domain of the bar. Clear negative gradients are observed on increasing distance away from the central regions, with strong asymmetries and plateau like features present.}
\label{fig:profiles}
\end{figure*}
In practice we compute the marginalisation over the nuisance parameters $(i, \theta)$ by summing the likelihood over a two dimensional Gaussian grid with mean $(i, \theta)$ values of $30^{\circ}$ and $235^{\circ}$. The covariance matrix of the Gaussian prior was forced to be diagonal with respective widths of $\sigma_{i} = 5^{\circ}$ and $\sigma_{\theta} = 10^{\circ}$. These choices reflect range of values reported in the literature; recently \citet{Choi_2018} inferred an LMC inclination and position angle of $(25.86^{\circ}, 149.23^{\circ})$ from photometric data alone whereas \citet{vdm_2014} infer viewing angles of $(34.0^{\circ}, 139.1^{\circ})$ from field proper motions and old stellar line of sight velocities (note in both these cases, the quoted position angle is in the usual astronomical convention). We perform the fit on all giants within a $12^{\circ}$ aperture of the LMC. We also exclude the most central giants within $3.5^{\circ}$ so as to avoid the metal-rich bar and focus on stars primarily tracing the LMC disc. In doing so, we also mitigate the fact that our metallicity predictions for the most metal-rich stars, which dominate centrally, incur a degree of bias ($\sim 0.1-0.2$ dex) in our regression model. We further limit our analysis to giants with $\left \langle \mathrm{D_{NN}} \right \rangle < 0.06$ to remove giants whose photometric colours lie in the domain in which our regression struggles to perform adequately. In doing so, we limit our sample to 196\,216 red giants. We recover a metallicity gradient of $-0.048 \pm 0.001$ dex/kpc, consistent with that of \citet{Choudhury_2016} who found a gradient of $-0.049 \pm 0.002$ dex/kpc in their analysis of LMC RGBs, as well as that of \citet{Cioni_2009} whose value of $-0.047 \pm 0.003$ dex/kpc was obtained from a sample of LMC AGB stars. For our intercept term we recover a value of $-0.656 \pm 0.004$. 

In Fig~\ref{fig:lmc_grad}, we show the radial metallicity profile of our LMC giants. Our fit describes the negative profile well through the LMC disc, with the inner regions being the most metal enhanced. We show the metallicity profiles along the projected bar major and minor axes for both Clouds in Fig.~\ref{fig:profiles}; the giants used in the figure were selected to lie between the hashed and dotted lines in Fig.~\ref{fig:lmc_smc_maps} respectively. For the LMC, we rotate into a coordinate system that is approximately bar aligned through a clockwise rotation of $30^{\circ}$ into the system we denote ($\textup{L}_{\textup{rot}}, \textup{B}_{\textup{rot}}$). In the Magellanic Stream coordinate system, the major axis of the SMC bar is very nearly aligned with the vertical and so no rotation was performed. Both of the Clouds display a flattening of their metallicity profiles centrally in the bar dominated regions. Outside of this domain, the profiles show clear negative gradients outwards into the disc. This is consistent with the findings of \citet{McKelvie_2019}, who studied a sample of 128 barred galaxies finding both the age and metallicity gradients to be flatter in the bar as opposed to the discs of the galaxies, indicative of bars being confined structures, efficient in radially mixing their stellar populations (see \citet{Siedel_2016} also). Through the LMC disc, complex structure is observed with asymmetric gradients and plateau features present. The black solid line in these panels shows our model projected into this coordinate system to highlight the large degree of asymmetric metallicity structure that exists throughout the LMC; the western portion of the disc (i.e. towards positive $\Delta$L$_{\textup{rot}}$) shows a depletion in metallicity with respect to the model whereas the northern and southern regions show mild excesses. Whilst we have hesitated to quantify the metallicity gradient in the SMC, owing to the uncertainty of the dwarf's morphology, Fig.~\ref{fig:profiles} demonstrates asymmetric gradients in the smaller Cloud also. 
\begin{figure*}
\centering
	\includegraphics[width=\textwidth]{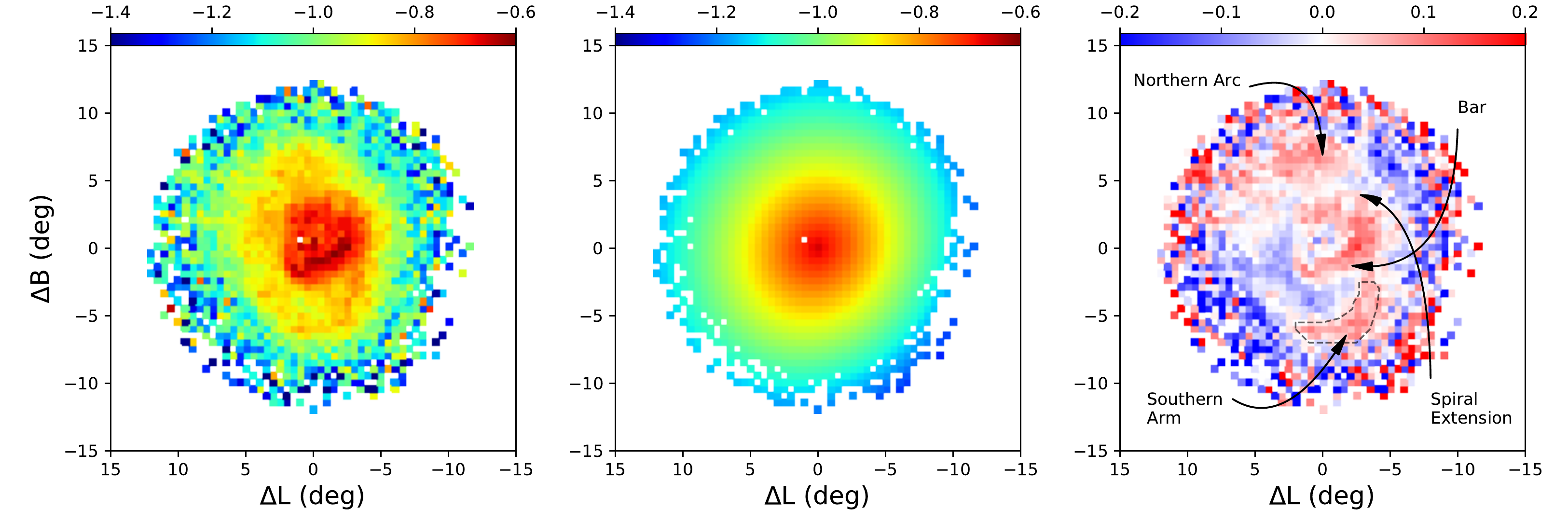}
    \caption[width=\columnwidth]{\textit{Left panel}: giants within $12^{\circ}$ of the LMC are shown in Magellanic Stream coordinates (relative to the LMC centre) with pixels coloured by mean metallicity. The metal-rich central region traces the bar and the dominant spiral arm of the Cloud. The northern portions of the disc trace arc like features and the southern region shows a spiral like feature with moderately enhanced metallicity. \textit{Middle panel}: Mean metallicity map of and inclined disc with viewing angles of \citet{Choi_2018} and radial metallicity function defined by the parameters found by out fitting method. \textit{Right panel}: We subtract the metallicity of the model pixels from that of our data to highlight regions that are enhanced/depleted with respect to a disc whose radial metallicity distribution follows the form of Eq.~\ref{eq:model}. Red pixels correspond to metal-richer regions and blue pixels metal-poorer with respect to the model. The most striking revelation in doing this is that of the LMC's metal-rich bar and inner northern spiral arm emanating from the north-west end of the bar. giants used to make this figure were subject to our nearest neighbour selection in feature space. Prominent regions, other than the bar, of enhanced metallicity are annotated and labelled.}
\label{fig:residual}
\end{figure*}

In Fig.~\ref{fig:residual}, we compare the mean metallicity maps of our LMC giants with that of an inclined disc whose radial metallicity profile follows that in our inference. The rightmost panel shows the metallicity residual, obtained by subtracting the model from the data, in which the LMC bar and main spiral arm are clearly revealed as centrally enhanced regions. The main spiral arm of the LMC is a feature usually only observed in young stellar tracers (ages $\lessapprox$ 1 Gyr) such as main sequence stars and supergiants (e.g. Cepheid Variables), notably so in the recent morphological mapping of the LMC by \citet{Youssoufi_2019} using VMC data. Utilising stellar synthesis models to calibrate stellar ages, they obtain age estimates for LMC stellar populations right across the CMD. Panels B and C of their Fig. 5 show the extent of the main spiral arm, with an additional faint arm emerging to the north of it; such a bifurcation is revealed in our metallicity residuals and is annotated in the figure as a spiral extension. In the northern regions of the outer disc we isolate an arc like area of metallicity enhancement. This portion of the LMC is coincident with the structure labelled "Arc" in Fig. 3 of \citet{Besla_2016}, lying $\sim 5-7^{\circ}$ above the LMC centre with no symmetric counterpart in the southern regions. 
\begin{figure}
\centering
	\includegraphics[width=\columnwidth]{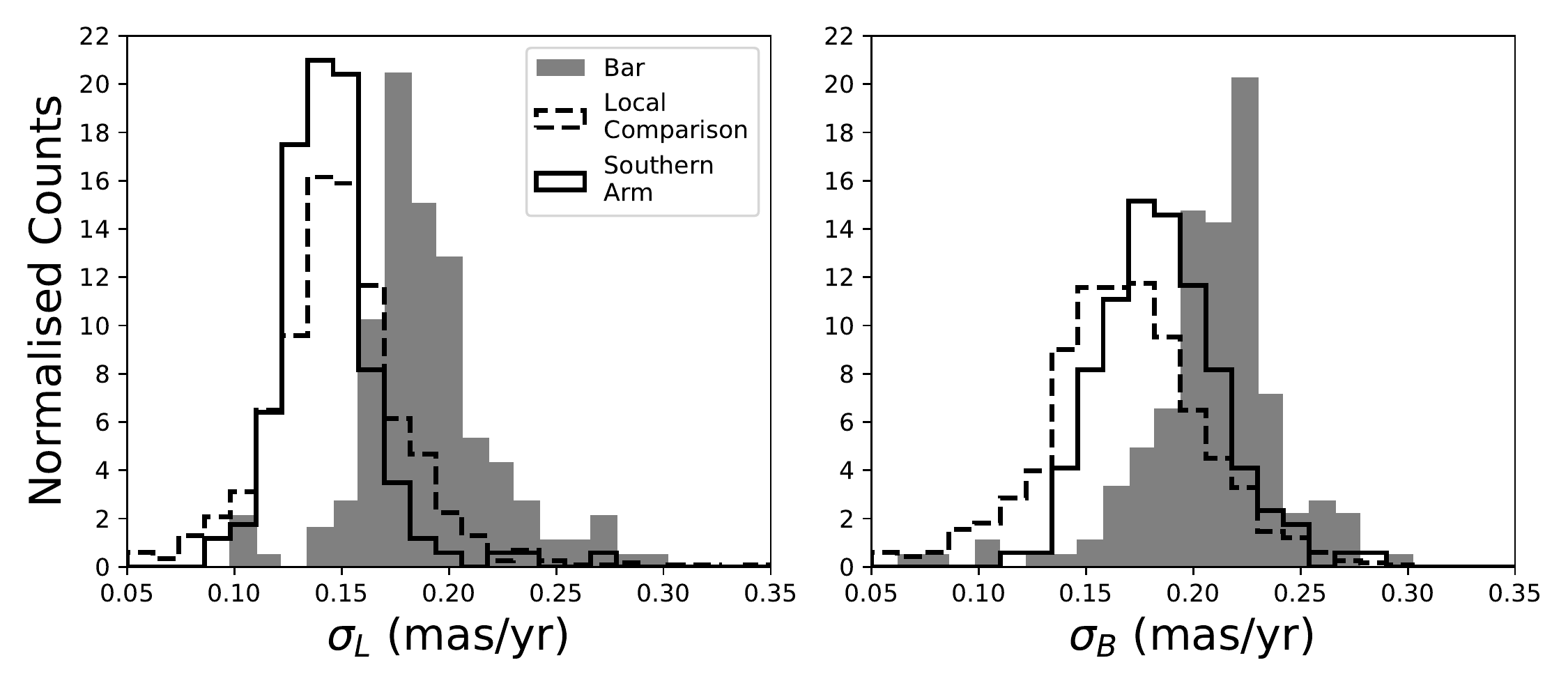}
    \caption[width=\columnwidth]{Histograms show reflex corrected proper motion dispersions in the Magellanic Stream coordinate system (L,B). The black lined histogram corresponds to pixels bound by the polygon in Fig.~\ref{fig:residual} and the dashed line histogram are stars bound by the aperture described in the text. The grey histogram represents the metal-rich pixels in the LMC bar region which are distinctly hotter than the two disc populations. Stars in the Southern Arm appear to share similar dispersions to those lying at a similar (projected) radius across all azimuthal angles.}
\label{fig:pm_disp}
\end{figure}
\begin{figure*}
\centering
	\includegraphics[width=\textwidth]{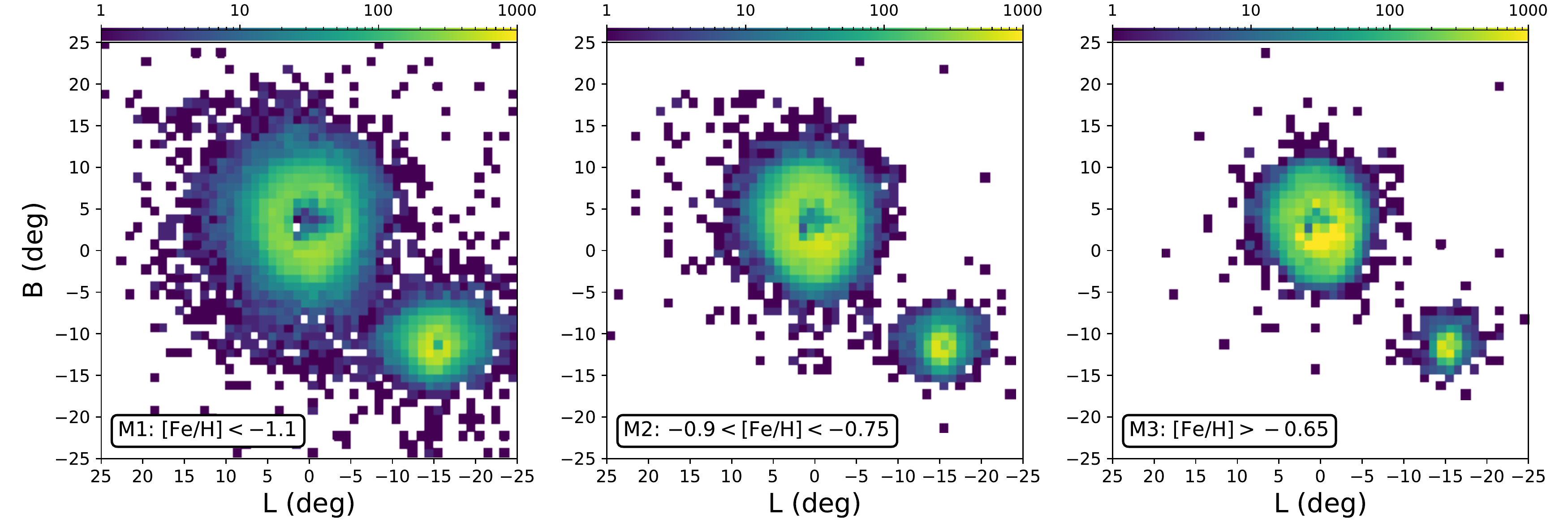}
    \caption[width=\columnwidth]{Stellar density maps of our red giants in the three metallicity bins described in the text with metal-poor in the leftmost panel and metal-rich in the right. The general evolution in morphology is a decrease in outer substructure in both Clouds on increasing stellar metallicity. The most metal-rich bin shows the LMC and SMC to be isolated, whereas the metal-poor bin shows the Clouds to be connected, both in the Magellanic Bridge region and further south where the outer stellar arc of the LMC attaches to the eastern edge of the SMC. The most metal-poor giants in the SMC shows a high central density with the outer regions appearing stretched horizontally, likely due to tidal interactions with the LMC. At the metal-rich end, the SMC is much more compact and shows a weaker signature of tidal disruption.}
\label{fig:density_map}
\end{figure*}
\begin{figure*}
\centering
	\includegraphics[width=\textwidth]{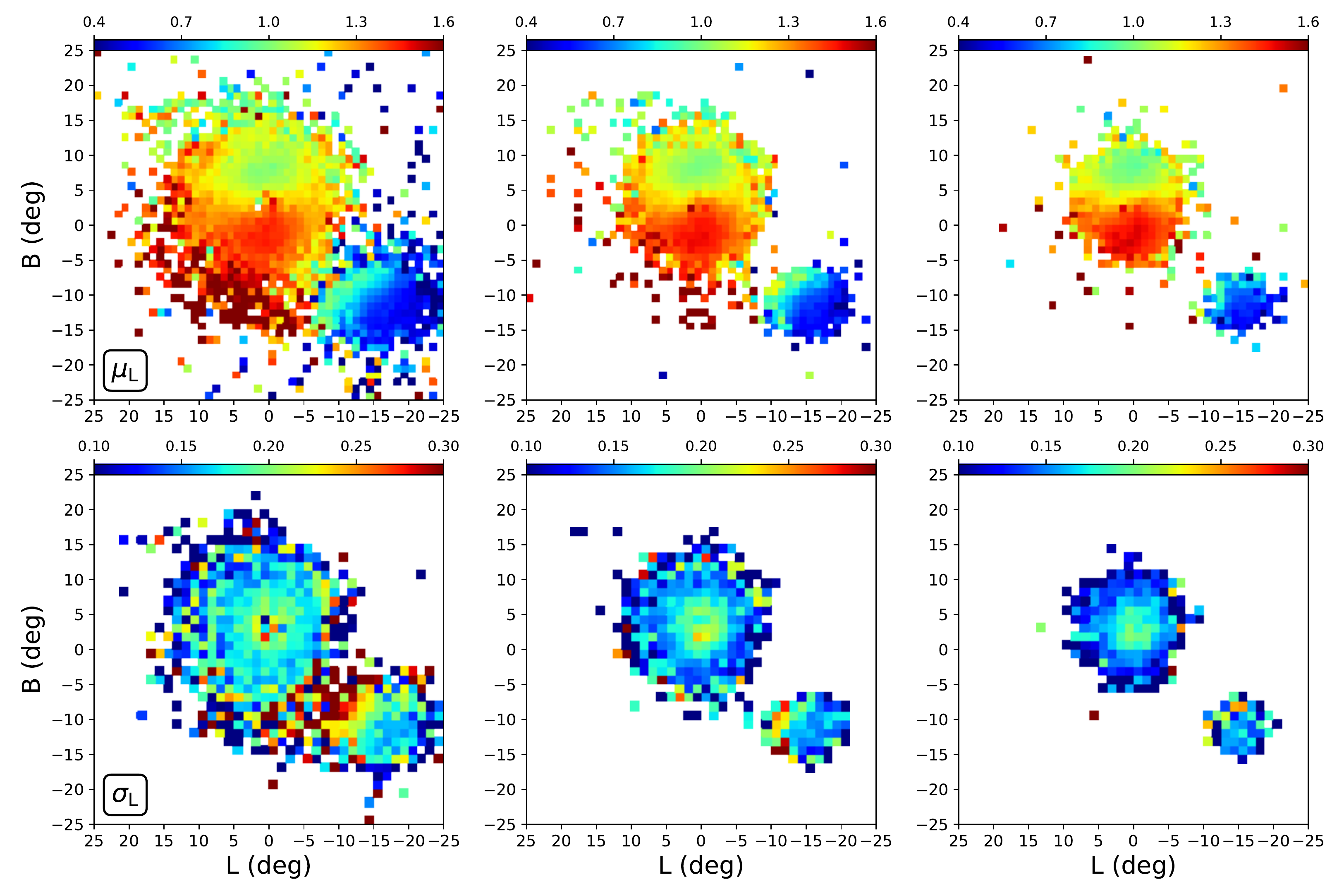}
    \caption[width=\columnwidth]{In the top panels, we show reflex corrected, mean $\mu_{\textup{L}}$ in three metallicity bins M1, M2 and M3 as in Fig.~\ref{fig:density_map}. The lower panels show the respective dispersion in each pixel. The LMC displays a coherent rotation signal across all the metallicity bins, with the two outer arms appearing to lag somewhat behind the inner disc. The SMC shows a gradient in its motion, with the edge nearest the LMC displaying prominent motion towards the larger Cloud. The greatest dispersion in the motion of the giants appears in the metal-poorest bin at the interface between the Clouds - a region where we expect mixing of stellar populations and turbulent motion due to tidal interactions within the system. The central-most region of the LMC, in the vicinity of the bar and spiral arm, is mildly visible as a region of enhanced dispersion.}
\label{fig:pmL}
\end{figure*}
This feature is likely a remnant of tidal interactions between the LMC and SMC with the metallicity enhancement observed in this outer region suggestive of metal-rich stars having been stripped outwards by the process. The curious spiral/arc like feature seen in Fig.~\ref{fig:lmc_smc_maps} is also revealed in Fig.~\ref{fig:residual}, which we label as a Southern Arm, wrapping from $\sim -4^{\circ}$ to $2^{\circ}$ in latitude and looks to emanate from the bar downwards by $\sim 6^{\circ}$. This residual feature is largely coincident with the ring like over-density uncovered by \citet{Choi_2018_b}. Through modelling the LMC stellar density with a disc and bar component, their residuals revealed a structure akin to ours composed of stars older than $\sim 1$ Gyr. They observe the feature to wrap around more extensively on the western side of the LMC in comparison to our metal-rich feature, which diminishes at ($\Delta$L, $\Delta$B) $\sim (2^{\circ}, -6^{\circ})$ in our residuals, and is more reminiscent of a spiral/arc as opposed to a ring. Curiously however, the signal of their over-density appears to diminish in strength near to the tip of our residual spiral. Beyond this, the \citet{Choi_2018_b} over-density becomes patchy as it continues to wrap around the LMC centre. Looking at our data in the left panel of Fig.~\ref{fig:residual}, we observe similar behaviour; the coherent spiral-like structure emerges from the south-west end of the bar and wraps eastward until ($\Delta$L, $\Delta$B) $\sim (2^{\circ}, -6^{\circ})$, beyond which we see a patchy continuation of slight $\feh$ enhancement. We attempt to determine if the Southern Arm stars are kinematically distinct in any way by considering their proper motion dispersions. To do so, we first select all pixels bound by the polygon in Fig.~\ref{fig:residual}. We further select all pixels within a $5-7.5^{\circ}$ aperture around the LMC centre, rejecting those within the polygon, to represent stars at a similar radius as the arm for a 'local' comparison. We also select metal-rich pixels associated with the LMC bar. As a simple investigation, we plot the histograms of pixel proper motion dispersion in Fig.~\ref{fig:pm_disp} where we see both disc populations to co-exist in proper motion dispersion, distinctly cooler than the bar region. This combined with the fact that this spiral like structure is a metal-rich feature is suggestive of it harbouring LMC disc stars. We conjecture that these stars may have been perturbed in some way, giving rise to the coherent, spiral like structure we see. Many barred galaxies display a symmetrical appearance, with spiral arms emerging from both ends of the bar. The LMC however has long been known to possess only one dominant spiral arm. Recently, \citet{Ruiz_Lara_2020} determine the LMC spiral arm to be a spatially coherent structure that has been in place for the last $\sim 2$ Gyr, supporting the notion that it was seeded through a historic close encounter with the SMC, as observed in simulations \citep[see e.g.][]{Besla_2012, Pearson_2018}. The N-body simulations of \citet{Berentzen_2003} demonstrate that the collision of a small companion with a larger barred galaxy can seed complex structures such as spiral arms, stellar rings and stellar spurs, and those of \citet{Walker_1996} further display the prominence of one sided spiral features seeded by the accretion of a satellite galaxy onto a large disc galaxy; single spiral arms emanate from the bar and wrapped spiral arms form in the disc. It is likely that the southern spur we observe has arisen from the historic interactions between the LMC and SMC and is a metal-rich counterpart of the structure observed by \citet{Choi_2018}. Our residual maps show the northern and southern portions of the LMC to be metal enhanced with respect to the eastern and western regions, an observation consistent with that of \citet{Choudhury_2016}, indicative of the former regions being most disrupted by previous tidal interactions. The region of slight metallicity depletion, running interior to the Southern Arm, is evident in the top left panel of Fig.~\ref{fig:profiles} as the plateauing region at $\Delta \textup{L}_{\textup{rot}} \sim 3^{\circ} - 5^{\circ}$, beyond which the metallicity diminishes as outer metal-poor stars begin to dominate. The two slight bumps in metallicity in the bottom left panel of Fig.~\ref{fig:profiles} correspond to the northern arc and Southern Arm features we present in Fig.~\ref{fig:residual}. 

\subsection{Slicing the Clouds by Metallicity}
\begin{figure*}
\centering
	\includegraphics[width=\textwidth]{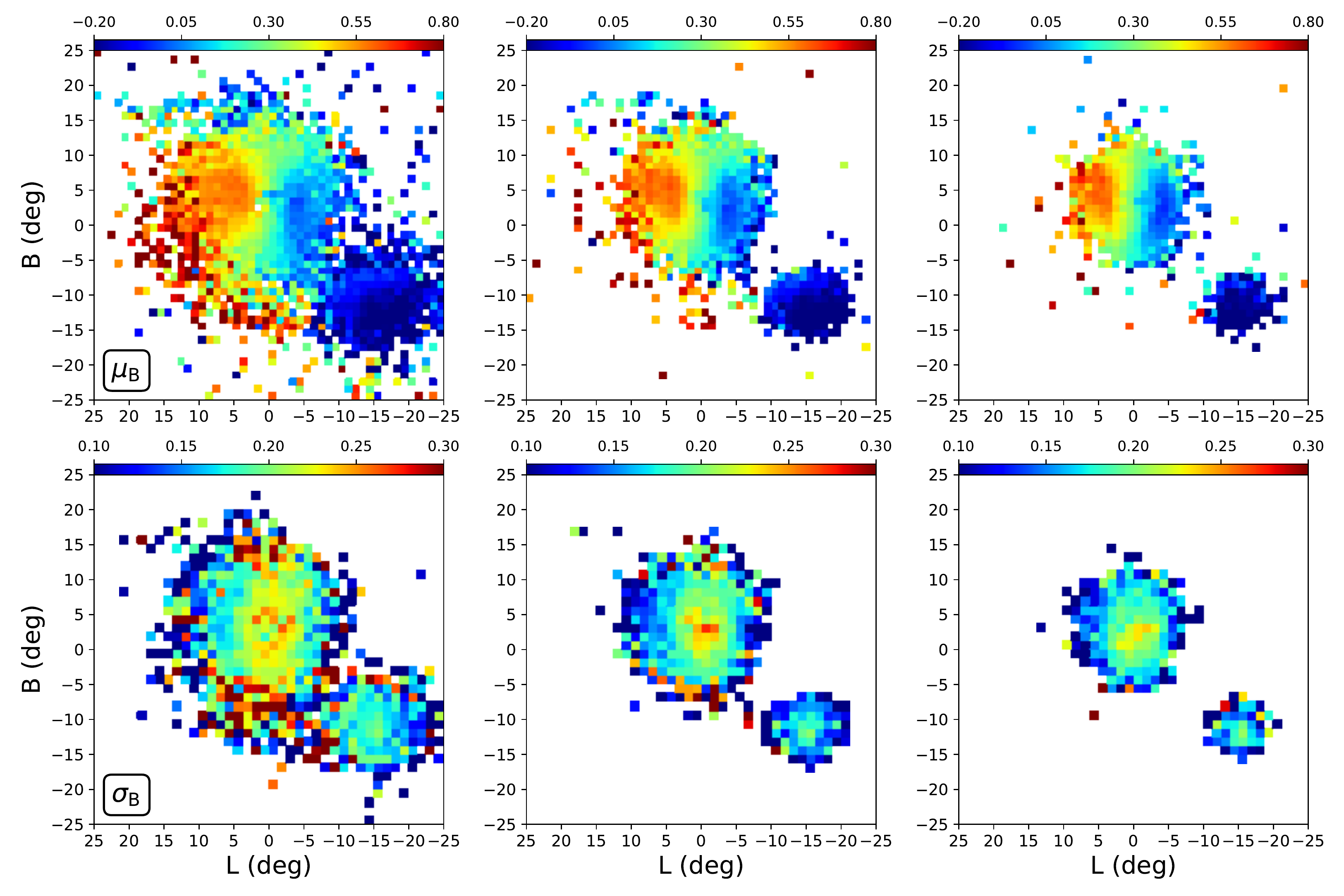}
    \caption[width=\columnwidth]{As in Fig.~\ref{fig:pmL} but pixels now coloured by mean $\mu_{\textup{B}}$ in the top panels and corresponding dispersions in the lower panels. Again the dominant signal within the LMC is that of ordered rotation within the disc. The regions of most turbulent motion appear to exists at the bases of the two outer spiral like features. Through the inner regions of the disc, there also appears to be enhanced dispersion in this direction in comparison to that of $\mu_{\textup{L}}$. This is perhaps indicative of the tidal disruption endured by the LMC occurring mostly in the north-south regions where the majority of stellar substructure is observed (see Fig.~\ref{fig:residual}).}
\label{fig:pmB}
\end{figure*}
We consider the broad scale morphology of the Clouds as a function of metallicity in Fig.~\ref{fig:density_map} where we show the stellar density in the three metallicity bins of: $\feh < -1.1,  -0.9 < \feh < -0.75$ and $\feh > -0.65$, each containing $\sim 44,000$ giants. We will refer to these bins as M1, M2 and M3 hereafter. Although there exists a degree of scatter between the bins due to the uncertainty in our metallicity predictions, we present this map to provide a sense of the general structural trends of the Clouds on increasing metallicity. The majority of outer substructure around the Clouds is made of the metal-poorer stars in M1. It is in this bin that the northern substructure identified by \citet{Mackey_2016} is most prominent. A region at the base of this structure was recently analysed by \citet{Cullinane_2020}, finding it to be kinematically perturbed from an equilibrium disc, likely through SMC/MW interactions. Furthermore, the southern stellar stream like feature that connects to the SMC, first observed by \citet{Clouds}, is only apparent in this bin and appears to be the symmetric counterpart to the northern outer arc. The LMC+SMC+MW N-body simulations conducted by \citet{Clouds} found that as recently as $\sim 150$ Myr after a close encounter with the SMC, such outer spiral-like features can be induced in the LMC. Indeed, the SMC is thought to have experienced a direct collision with the LMC on a timescale comparable to this from the kinematical modelling of \citet{Zivick_2018}. Another diminishing feature on increasing metallicity is the density of stars in the old stellar Magellanic Bridge region, likely composed of outer, tidally stripped stars originating from both the LMC and SMC. The general shape of both Clouds also appears to evolve in a sensible way, presenting themselves as extended, rather diffuse objects at the metal-poorer end in M1 through to much more centrally concentrated objects at the metal-rich end in M3. In the case of the SMC, it demonstrates a degree of ellipticity in M1 (and slightly so in M2), indicative of strong tidal disruption induced in the outer metal-poor regions. 

We consider the proper motions of our giants in the three bins in Fig.~\ref{fig:pmL} and Fig.~\ref{fig:pmB}. We correct for the solar reflex motion assuming a constant heliocentric distance of 49.9 kpc (with the main focus being the LMC). The dominant signal is that of rotation within the LMC disc, apparent by the gradient across the Cloud. As in \citet{Clouds}, the northern and southern arm like features display a coherent rotation signal, lagging that of the LMC disc; both arms bear motions that are consistent with the bulk of the LMC and appear distinct from the proper motions of the SMC. In Fig.~\ref{fig:pmL}, a significant portion of the SMC nearest to the LMC shows prominent motion towards the larger Cloud, with the signal appearing to persist across the three metallicity bins. The lower left panel of Fig.~\ref{fig:pmL} shows there to be significant dispersion in M1, precisely at the SMC edge of enhanced proper motion and at the LMC-SMC interface, where a mixture of Cloud populations is to be expected. The SMC is known to be disrupting \citep[see e.g.][]{Zivick_2018, DeLeo_2020} and the perturbed motions we observe here are likely a result of this, with the LMC violently dragging the eastern edge of the SMC towards it. The fact that we observe this signature in all three of the top panels indicates that the disruption is severe, penetrating through to the more centrally concentrated metal-rich giants in M3. With respect to the motions of $\mu_{\textup{B}}$, the two main regions of significant heating appear in M1: at the base of the northern arm and another diametrically opposed to this in the southern region. Between these two points we also observe a vertical region of moderate dispersion in the LMC. That is, the dispersion in $\mu_{\textup{B}}$ is greater along north-south direction than so along east-west. This may be a consequence of how the LMC has been perturbed by the SMC, as the majority of substructure observed in Section.~\ref{sec:grads} is observed to lie in the northern and southern regions of the LMC disc. In each bin, the central bar region also displays a degree of enhanced dispersion, both in $\mu_{\textup{L}}$ and $\mu_{\textup{B}}$, owing to the complex orbits hosted by galactic bars. 
\subsection{SMC shape and disruption}
\label{sec:smc_shape}
\begin{figure}
\centering
	\includegraphics[width=\columnwidth]{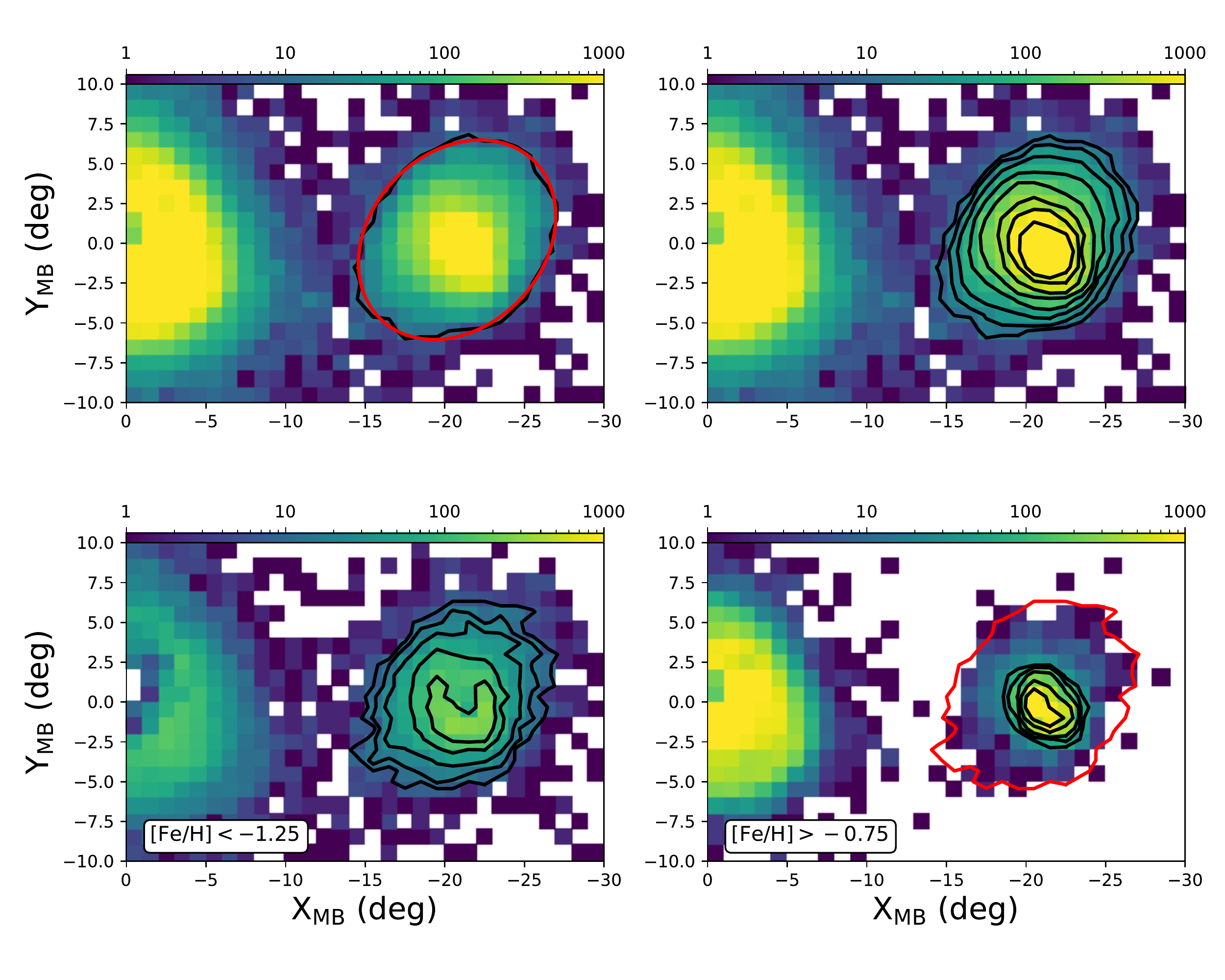}
    \caption[width=\columnwidth]{We show the stellar density of stars near to the SMC in Magellanic Bridge coordinates. \textit{Top left}: Black solid line is an outer isodensity contour corresponding to 10 stars per square degree. We fit an ellipse to this contour, shown by the solid red line, whose ellipticity is mild at $\sim 0.21$. \textit{Top right}: We overlay logarithmically spaced isodensity contours on the SMC where a clear S-shape is apparent, reminiscent of tidal tails and consistent with the RR Lyrae SMC morphology observed by \citet{Vasily_2017}. \textit{Bottom left}: We now only show giants with predicted metallicities of $\feh < -1.25$ dex and contours trace pixels corresponding to the $2.5^{\textup{th}}, 5^{\textup{th}}, 10^{\textup{th}}, 25^{\textup{th}}$ and $50^{\textup{th}}$ percent contour levels for stars around the SMC. \textit{Bottom right}: We only consider stars with $\feh > -0.75$ dex and trace the same contour levels with black solid lines. The red line corresponds to the outermost contour of the metal-poor SMC giants. Whilst the metal-rich giants are located much more centrally within the SMC, the outer regions still bear the signature of tidal disruption, indicating that the SMC has been significantly disrupted across a range of stellar populations. }
\label{fig:S_shape}
\end{figure}
In Fig.~\ref{fig:S_shape}, we show the density of stars lying around the SMC in Magellanic Bridge coordinates. The top left panel of the figure shows an ellipse fitted to an outer SMC density contour for which a mild ellipticity value of $\sim 0.21$ is obtained. Upon plotting a set of iso-density contours to the SMC giants in the top right panel, a distinctive twisting is seen, with the outer regions displaying an S-shape that is characteristic of a tidally perturbed system. Similar SMC morphology was observed by \citet{Vasily_2017} in the $Gaia$ DR1 data. They found that the orientation of the S-shaped tails aligned conspicuously with the SMC's proper motions vector (relative to the LMC). Based on this, they designated the tail nearest to the LMC to be the trailing arm and that stretching towards the top right in Fig.~\ref{fig:S_shape} to be leading. In the lower panels of the figure, we have sliced the stars into two metallicity bins, symmetrically offset from the mean of our SMC metallicity distribution, so as to investigate the morphology of a relatively metal-poorer and metal-richer subset of SMC giants. The bottom left panel of Fig.~\ref{fig:S_shape} shows the stellar density of stars with $\feh < -1.25$ dex. These metal-poorer giants constitute a relatively diffuse, fluffy population with their outermost contour again showing the tidally symptomatic S-shape. The bottom right panel of the figure shows stars with predicted metallicities $\feh > -0.75$ dex and we show the same iso-density contours as in the metal-poor bin in black for comparison. The bulk of these metal-richer giants are confined to the most central regions, with little extent beyond $\sim 5^{\circ}$ of the Cloud centre. The red solid line traces the outer-most contour level of the metal-poorer bin which highlights that the outskirts of these relatively metal-richer giants also show symptoms of tidal disruption.

\begin{figure}
\centering
	\includegraphics[width=\columnwidth]{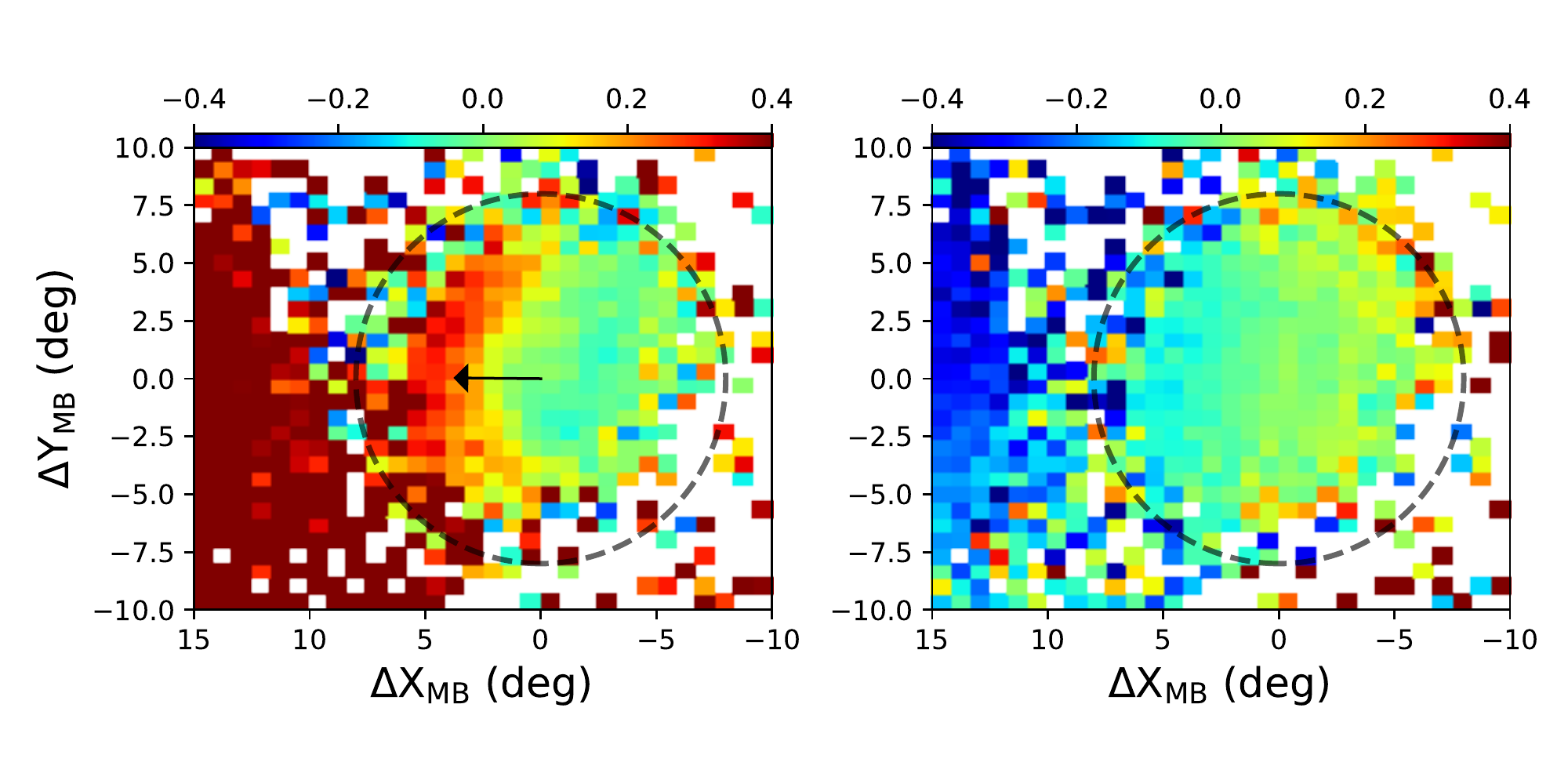}
    \caption[width=\columnwidth]{SMC giants shown in the Magellanic Bridge coordinate system. Pixels in the left (right) panel are coloured by mean proper motion along $\textup{X}_{\textup{MB}}$ $\textup{Y}_{\textup{MB}}$. The black arrow in the left panel points towards the centre of the LMC in this system. There is little sign of internal motion in $\mu_{\textup{Y}_{\textup{MB}}}$ with a distinct lack of rotation signal. In the left panel, a clear gradient is seen with SMC giants residing on the side nearest the LMC clearly being disrupted toward the larger Cloud.}
\label{fig:smc_motion}
\end{figure}

We consider the motion of the SMC in a simple fashion by first studying the spatially averaged proper motion components in Fig.~\ref{fig:smc_motion}. In the figure, we have set the origin of the coordinate system to the SMC centre. The pixels in the left (right) panel are coloured by mean $\mu_{\textup{X}_{\textup{MB}}}$ ($\mu_{\textup{Y}_{\textup{MB}}}$), where we have centered the proper motion distributions around the bulk values for our SMC giants. We compute these based on the mean proper motions of giants lying within a $3^{\circ}$ aperture of the SMC for which we obtain the SMC motion to be $(\mu_{\alpha}, \mu_{\delta}) = (0.65, -1.21) \, \textup{mas yr}^{-1}$, consistent with the recent determination of \cite{DeLeo_2020} as well as that of \citet{Zivick_2018}. We have also corrected the proper motions for the perspective expansion/contraction induced by the systematic centre of mass motion along the line of sight, following \citet{vandeVen_2006}. The left panel of the figure shows stars residing on the edge nearest the LMC (on the side of the trailing tail), being pulled directly towards the larger Cloud, with giants on the opposite leading edge of the dwarf showing little sign of such effects. Further, the motion of the giants in $\mu_{\textup{Y}_{\textup{MB}}}$ shows very little structure, with no indication of ordered rotation. Note that the faint vertical, banding structure seen in this panel is an artefact of the $Gaia$ scanning law (see \citet{Helmi_2018}). Thus, in the picture we present here, it would seem that the region of the SMC nearest the LMC, and coincident with the trailing arm, is being violently hauled towards the LMC. Recent detailed kinematic analysis of SMC RGB stars by \citet{DeLeo_2020} found the dwarf to be undergoing strong tidal disruption, with a net outward motion of RGB stars in the direction of the LMC. The giants display strong tangential anisotropy in their proper motion dispersions, right down to the SMC centre. Through comparison to a suite of N-body simulations of the Clouds in orbit about the MW, they argued this effect is due to unbound material lying in front of the SMC, distinct in their kinematics due to tidal stripping by the LMC. The proper motion analysis of \citet{Zivick_2018} also showed the SMC to bear little sign of ordered rotation, but rather mean ordered motion radially away from the galaxy in its outer regions, consistent with heavy disruption. 
\begin{figure*}
\centering
	\includegraphics[width=0.9\textwidth]{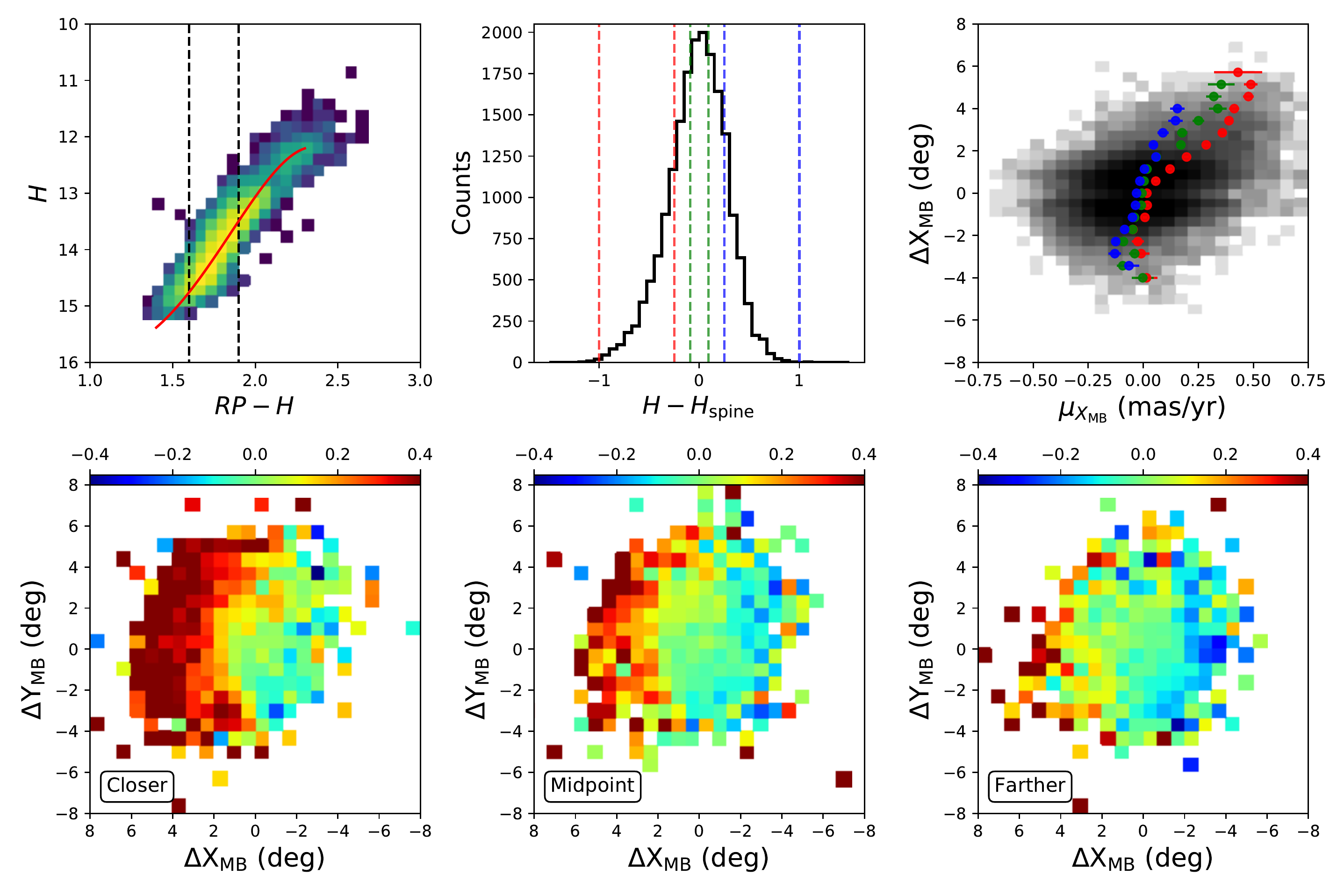}
    \caption[width=\textwidth]{\textit{Upper left}: CMD of SMC giants within a fixed metallicity range of $-1 < \feh < -0.9$ dex. The solid red line shows the polynomial fitted to the RGB in this bin, offsets from which are taken as a proxy for heliocentric distance. As described in the text, we have fitted such polynomials to SMC giants lying in the range $-1.2 < \feh < -0.5$ dex, over metallicity bins of width 0.1 dex. In computing the magnitude offset of our giants, we only consider those falling between the black hashed lines in the panel. \textit{Upper middle}: We show the distribution of the magnitude offset of the SMC giants we select to analyse. The distribution is centered about zero and appears to be relatively symmetric about its peak, with a slight tail to negative values with these stars likely lying in front of the dwarf. Three bins in this distance proxy where selected to study the behaviour of giants lying in front of, coincident with and behind the centre of the SMC. The boundaries of these bins are shown by the red, green and blue hashed lines respectively. \textit{Upper right}: Grey scale pixels show the logarithmic density of stars in their relative offset from the SMC centre as a function of their motion in $\mu_{\textup{X}_{\textup{MB}}}$. There is a distinct population of stars with excessive motion in $\mu_{\textup{X}_{\textup{MB}}}$ along increasing $\Delta \textup{X}_{\textup{MB}}$, towards the LMC. The scatter points are coloured by their respective distances, and show mean proper motion values as a function of $\Delta \textup{X}_{\textup{MB}}$. We see that it is those giants lying predominately in front of the SMC who exhibit such perturbed motion. This sequence becomes distinct in its behaviour very centrally, and rapidly evolves out to high values of $\mu_{\textup{X}_{\textup{MB}}}$ on increasing angular distance away from the SMC core. Error bars represent the standard error weighted by the Poisson noise in each bin. \textit{Lower panels}: We show the SMC giants in the three distance bins, with the closet in the left panel and the most distant in the right. Pixel colours correspond to mean values of $\mu_{\textup{X}_{\textup{MB}}}$. This view of the SMC reinforces the result of the upper right panel, where we can clearly see the giants lying closer to use being those most heavily disrupted in the direction of the LMC.}
\label{fig:smc_distances}
\end{figure*}
The geometry of the SMC is complex, with a substantial line of sight depth. The north-eastern regions of the dwarf appear to lie closer to us than its western edge, as perceived through numerous stellar tracers including star clusters, red clump stars, Cepheid Variables and RR Lyrae \citep[see e.g.][]{Crowl_2001, Haschke_2012, Subramanian_2012, Deb_2015, Scowcroft_2016, Muraveva_2018, Deb_2019}. The SMC red clump analysis of \citet{Nidever_2013} revealed a stellar structure lying eastward at $\sim 10$ kpc in front of the dwarf, a structure they interpret to have been tidally stripped and dragged toward the LMC/Magellanic Bridge region. Further to this, \citet{Subramanian_2017} also found evidence for such a tidally stripped stellar structure in identifying a population of VMC red clump stars lying $\sim 12$ kpc in front of the main SMC body, tracing them from the direction of the Magellanic Bridge right down to $\sim 2.5^{\circ}-4^{\circ}$ of the SMC centre. Very recently, \citet{Omkumar_2020} identified this foreground red clump population in $Gaia$, tracing it from the inner $2.5^{\circ}$ out to $\sim 5^{\circ}-6^{\circ}$ from the centre of the SMC. We attempt to discern whether those giants whose on-sky motion is prominently towards the LMC in Fig.~\ref{fig:smc_motion}, are indeed lying closer to us as a result of tidal stripping by the larger Cloud. To do so we first select all SMC giants lying within the black hashed circle in Fig.~\ref{fig:smc_motion}. We then attempt to use each star's relative position within the observed CMD as a proxy for heliocentric distance. We first select stars for which we have predicted metallicities in the range of $-1.2 < \feh < -0.5$ dex, a region where there is sufficiently low bias and variance in our predictions that our estimates are reasonable. We then divide these giants into metallicity bins of width 0.2 dex and, considering each metallicity bin independently, fit a polynomial to the CMD of the giants, an example of which is shown by the solid red line in the upper left panel of Fig.~\ref{fig:smc_distances}. We do this to minimise the broadening of the CMD due to the spread in metallicities of the SMC population. Utilising these polynomials fits as base 'spines', we compute the magnitude offset $H-H_{\textup{spine}}$ for the selected SMC stars relative to the spine, such that stars with a negative (positive) offset are likely closer (farther) in distance. We only use stars which fall in the range $1.6 < RP-H < 1.9$, a region where the polynomial fits appear most reasonable and indicated in Fig.~\ref{fig:smc_distances} by the vertical black hashed lines. The upper middle panel of Fig.~\ref{fig:smc_distances} shows the distribution of this magnitude offset, where we see it to be centred at zero and approximately symmetric. The histogram shows a slight skew towards negative values, indicative of a higher relative proportion of SMC giants lying in front of the dwarf's core. We then choose three bins in $H-H_{\textup{spine}}$ initially centered at -0.3 mag, 0 mag and 0.3 mag, corresponding to $\sim \pm 10$ kpc in front of and behind the SMC, with the bin edges shown by the red, green and blue lines respectively. We have then allowed the widths of the two outer bins to be broad so as to roughly equalise the counts per bin (3000) and to encompass stars lying in the tails of the distribution. The upper right panel of the figure shows the logarithmic density of our selected SMC stars' spatial coordinate $\textup{X}_{\textup{MB}}$ as function of their motion in $\mu_{\textup{X}_{\textup{MB}}}$ (approximately) towards the LMC. From this density plot alone, there is a clear population of stars with high $\mu_{\textup{X}_{\textup{MB}}}$ values, deviating from the bulk beyond $\mu_{\textup{X}_{\textup{MB}}} \sim 2^{\circ}$. For the three distance bins we have defined, we show the mean proper motion binned over coordinate $\textup{X}_{\textup{MB}}$ where a clear distinction is observed between the different distance bins. At a value of $\textup{X}_{\textup{MB}} \sim 2^{\circ}$, the stars most likely lying in front of the SMC display a sharp turn toward stronger motion in the LMC direction, dominating the population of stars exhibiting high values of $\mu_{\textup{X}_{\textup{MB}}}$. It therefore seems that the stars apparently being ripped from the SMC are predominately those lying closer to us, tracing the disruption of the trailing arm down to the inner regions of the SMC. On increasing heliocentric distance, the signal diminishes with little sign of a companion leading arm in the proper motions. The bottom panels of the figure are coloured by mean $\mu_{\textup{X}_{\textup{MB}}}$ where we show each of the three distance bins independently, with stars lying in front (behind) of the SMC in the left (right) panel and stars approximately at the SMC distance in the middle panel. The evolution of disruption on increasing distance is evident, with the closer stars being those most strongly dragged towards the LMC. Thus, our interpretation is that the portion of the SMC in which we see strong motion towards the LMC in Fig.~\ref{fig:pmL} and Fig.~\ref{fig:smc_motion} consists largely of tidally stripped stars; stars that have been dragged outwards to closer heliocentric distances with their motions disrupted towards the larger Cloud, in effect forming a tidal tail trailing the SMC. We observe the signature down to angular separations of $\sim 2-3^{\circ}$ from the SMC centre, indicative of heavy disruption of SMC stars lying along our line of sight. The fact that we do not observe a significant kinematic counterpart to the trailing tail lying beyond the SMC (i.e the leading arm) is somewhat at odds with simulation models of the Magellanic system. Whilst the majority of simulations in the literature aim to primarily trace the gaseous features of the Clouds, those of \cite{Diaz_2012} do well in replicating the general properties of the Magellanic bridge region, with signs of stars being drawn eastward and towards the LMC \citep[see][for discussion]{Nidever_2013}. However, these models also predict a 'counter-bridge', lying at large heliocentric distances and predominately directly behind the SMC \citep[see also][]{Vasily_2017}. Such a feature is predicted to be much more diffuse than its bridge counterpart which may indicate why no stellar detection has been made, either in this work or by \citet{Nidever_2013} and \citet{Subramanian_2017}. Indeed, the counter bridge, if present, may be so diffuse that it exists as a purely gaseous feature with minimal stellar counterpart (see \citet{Diaz_2012}). We note that the stellar content of the SMC is complex, with stars spanning a range of metallicities and ages. In our above analysis, we have alleviated the effect that metallicity has in broadening the CMD. We cannot, however, account for the fact that giants of differing ages, at fixed metallicity and distance, will induce a magnitude offset in  $H-H_{\textup{spine}}$. Thus our distance proxy is rather a convolution of heliocentric distance, stellar age as well as reddening effects along the line of sight. Nonetheless, the results are tantalising, and appear in agreement with previous studies of SMC debris. As a consistency check, we applied our procedure to subsamples of LMC giants lying in the north-east and south-west regions of the disc. Given the viewing angles of the LMC, the north-east disc is lying at a closer heliocentric distance to us than the south-western edge. Thus, this should be detectable in our method; indeed we find a median magnitude offset of -0.12 mag for the giants located in the north-east and one of 0.03 mag for those in the south-west. These offset values are comparable to those of \citet{Li_2016} who applied a similar method to M-giants in the LMC. 

\subsection{Old stellar Bridge}
\label{sec:bridge}
\begin{figure}
\centering
	\includegraphics[width=0.9\columnwidth]{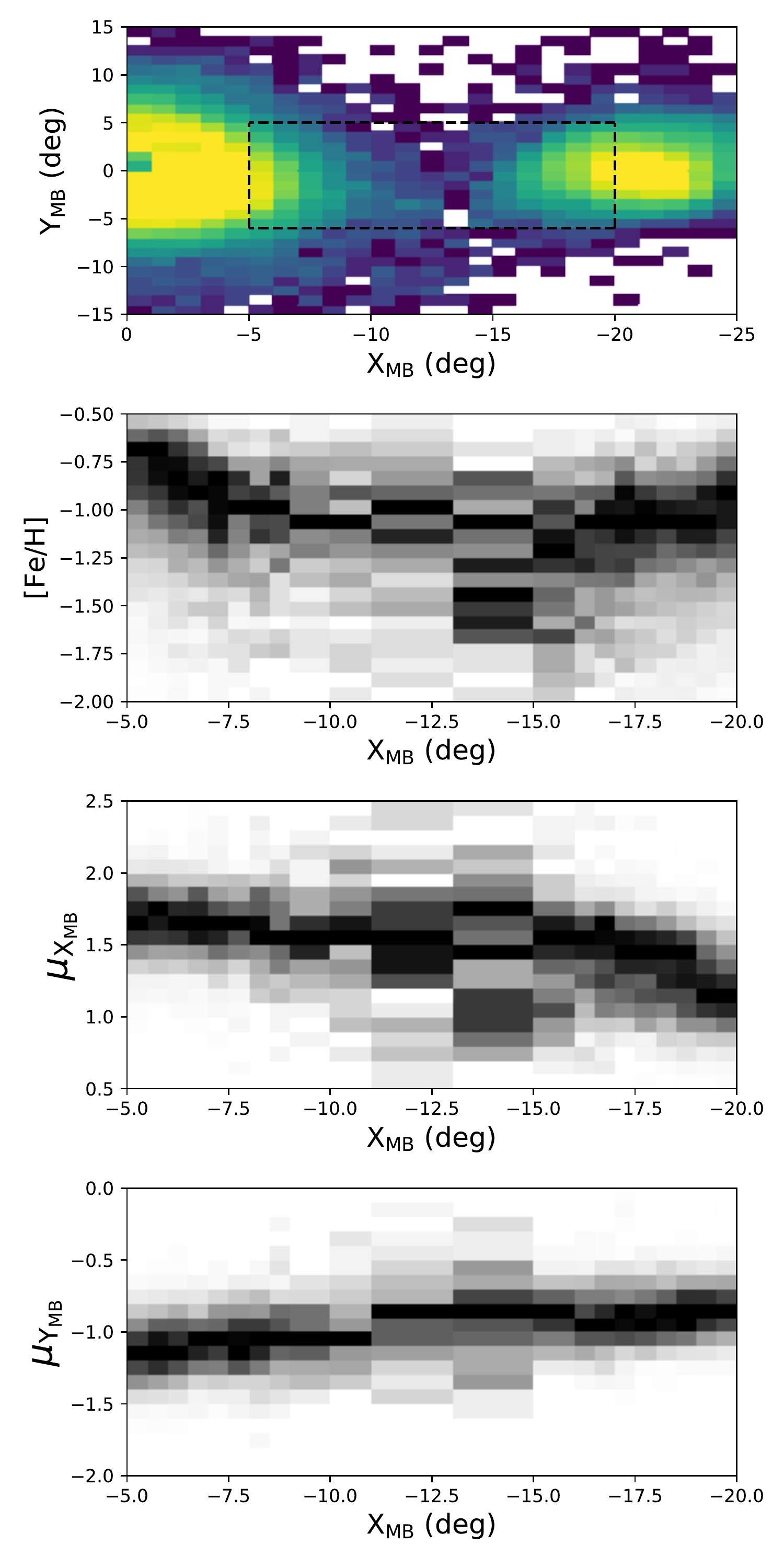}
    \caption[width=\columnwidth]{The top panels shows the logarithmic density of giants in Magellanic Bridge coordinates. In the region between the Clouds, there exits a population of stars whose nature we would like to probe. On selecting stars falling within the black rectangle, we show their metallicity along the bridge in the panel below. The column normalised density shows the evolution of giant metallicity upon the passage from the SMC to the LMC. Between the Clouds, it appears as if they are joined by a constant ridge at $\sim -1$ dex, a value consistent with both the outer regions of the LMC and the median value of all SMC giants. On moving outwards from the SMC, there appears to be two sequences: the ridge that connects to the LMC and a metal-poorer population that extends out to $\textup{X}_{\textup{MB}} \sim -13^{\circ}$. The lower two panels show the proper motions of the stars through the bridge region, with a relatively smooth transition between the Clouds being apparent. The motion along $\mu_{\textup{X}_{\textup{MB}}}$ shows a degree of dispersion that is consistent with Fig.~\ref{fig:pmL}.}
\label{fig:giant_bridge}
\end{figure}
Given that we have found evidence for a population of stars within $8^{\circ}$ of the SMC centre, lying in front of the dwarf, and displaying a kinematic signature of tidal disruption, a natural avenue of exploration is to now assess the stars lying directly in the inter-Cloud region as can be gleaned from Fig.~\ref{fig:S_shape}. Historically, the detection of a continuous path between the Clouds in intermediate and old stellar populations has proved elusive. Utilising 2MASS and WISE photometry, \citet{Bagheri_2013} selected a sample of giants, with ages ranging from $\sim 400 \, \textup{Myr} - 5 \, \textup{Gyr}$ in the bridge region, at low stellar density. The analysis of OGLE RGB and RC stars by \citet{Skowron_2014} found little evidence of a coherent stellar bridge, but rather posited that the presence of evolved stellar populations in this region stems from overlap of the MC's halos. They do however observe the diffuse structure of giants to lie predominately south of the young bridge. This is consistent with the findings of \citet{Vasily_2017} who, on selecting RR Lyrae out of the $Gaia$ DR1 catalogue, were able to trace a continuous structure between the Clouds, offset by $\sim 5^{\circ}$ south of the young Magellanic Bridge. They argued this offset was formed under the scenario that both the gas and old RR Lyrae were stripped coevally from the SMC. The stripped gas was accosted by the hot gas of the Milky Way corona in the form of ram pressure, essentially pushing it back and forming the offset. The spectroscopic study of inter-Cloud red giants by \citet{Carrera_2017} found the chemistry and kinematics of their intermediate aged sample to be consistent with that of an tidally stripped SMC population, adding weight to the notion that an older stellar bridge exists in some form. Subsequently, the deep DECam imaging of \citet{Mackey_2018} and the highly-pure giant sample of \citet{Clouds} revealed the presence of a tangled mix of ancient stellar populations in the Old Magellanic Bridge region confirming the earlier discovery of \citet{Vasily_2017}.
\begin{figure*}
\centering
	\includegraphics[width=\textwidth]{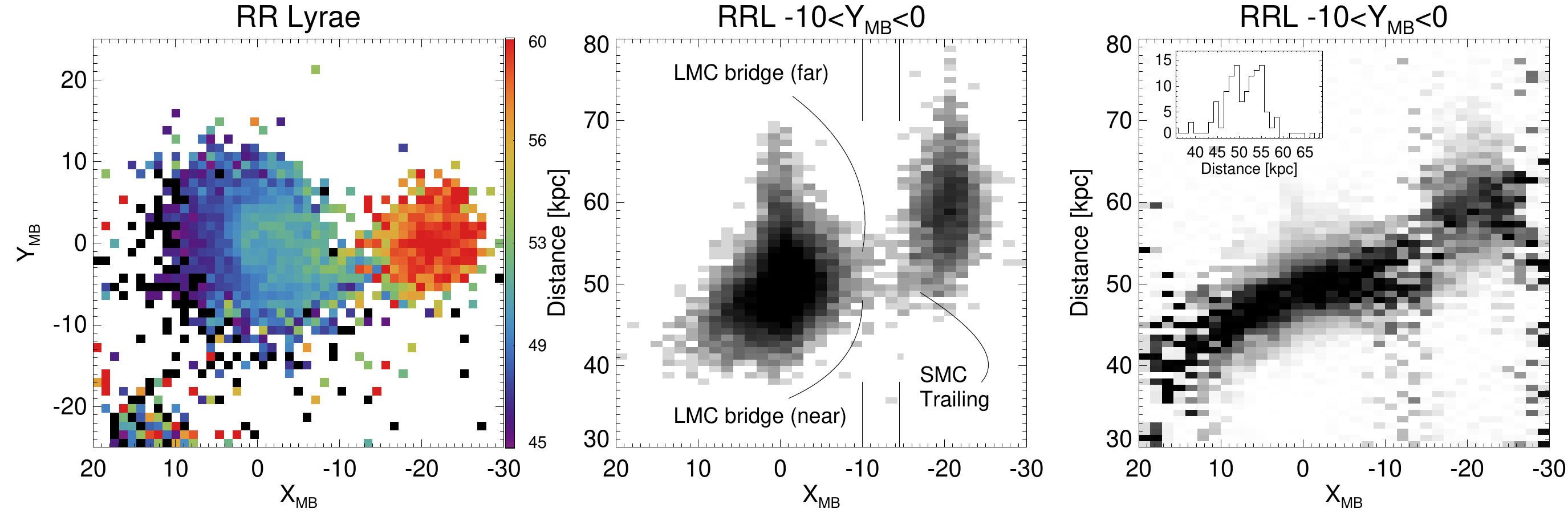}
    \caption[width=\textwidth]{We show the $Gaia$ DR2 RR Lyrae whose selection was described in Sec.~\ref{sec:data}. \textit{Left}: Pixels are coloured by mean RRL heliocentric distance. A gradient is observed across the face of the LMC disc owing to its inclination. We also see RRL residing in the eastern portion of the SMC to lie at closer heliocentric distances. Between the Clouds, we see the old stellar bridge with stars lying at distances consistent with both Clouds. \textit{Middle}: RRL distance is shown along $\textup{X}_{\textup{MB}}$ for stars lying in the slice of $-10^{\circ} < \textup{Y}_{\textup{MB}} < 0^{\circ}$. The SMC morphology is cigar-like, stretching along our line of sight. There appears to be two distinct structures lying between the Clouds at heliocentric distances of $\sim 50$ kpc and $\sim 55$ kpc. The inset panel shows the distance distribution for RRL lying in the slice $-14^{\circ} < \textup{X}_{\textup{MB}} < -10^{\circ}$ in which we observe a bi-modality, with the two peaks associated with the dual structures stretching between the Clouds. 
    }
\label{fig:RRL}
\end{figure*}

In the top panel of Fig.~\ref{fig:giant_bridge}, we show the stellar density of our giants in MB coordinates, with the black dashed box indicating the selection of stars lying in the bridge region. In the panel below, we show the distribution of metallicity of these giants progressively through the bridge. On moving through the outer LMC disc, the metallicity shows a steady decline, with the SMC profile remaining relatively flat out to $\textup{X}_{\textup{MB}} \sim -15^{\circ}$, just beyond which appears to lie a metal-poor sample of stars with $\feh \sim 1.5$ dex. Moving through the region of $-15^{\circ} < \textup{X}_{\textup{MB}} < -10^{\circ}$, the metallicity structure becomes more complex and difficult to discern owing to the low stellar density in this region. Interestingly however, there does appear to be a continuous ridge of constant metallicity, $\sim -1$ dex, running through this region, connecting the outer edge of the LMC out towards $\textup{X}_{\textup{MB}} \sim -15^{\circ}$. It is noteworthy that the metallicity along this ridge is consistent with the mean value of SMC giant's lying within the inner $4^{\circ}$ of the dwarf, as well as that of the outer regions of the LMC. The lower two panels of the figure trace the proper motions through the bridge region, with $\mu_{\textup{Y}_{\textup{MB}}}$ showing a relatively coherent structure that smoothly links the motions of the SMC to the LMC. The motion in $\mu_{\textup{X}_{\textup{MB}}}$ is less well defined through the bridge region, with greater scatter about the general trend. This is consistent with the high level of dispersion we see in the lower left panel of Fig.~\ref{fig:pmL} at the interface of the two Clouds, as well as our results from Section~\ref{sec:smc_shape}. The signature of a flow of stars through the bridge region is consistent with the findings of \citet{Zivick_2019} whose proper motion analysis of red giants through the bridge closely resembles our findings (see their Fig. 8). The relatively complex metallicity distribution in the inter-Cloud region, alongside the high dispersion in $\mu_{\textup{X}_{\textup{MB}}}$ render it difficult to exactly discern the origin of the stars lying in this region. It is highly likely however that the population of stars in this region are a mixture of both LMC and SMC giants. The origin of LMC debris in the bridge could be attributed to the effect of Milky Way tides acting on the larger dwarf, with the N-body simulation of \citet{Vasily_2017} and \citet{Clouds} demonstrating that such an effect can quite easily strip LMC debris to align well with the old stellar bridge. 

In Fig.~\ref{fig:RRL}, we show the sample of $Gaia$ DR2 RRL described in Sec.~\ref{sec:data} for which we are afforded heliocentric distance estimates. The left panel of the figure shows the Clouds in MB coordinates with pixels coloured by mean heliocentric distance, where we see RRL lying at distances consistent with both the LMC and SMC in the old stellar bridge. Selecting RRL in the slice of $-10^{\circ} < \textup{Y}_{\textup{MB}} < 0^{\circ}$, we show the distance distribution of these stars along $\textup{X}_{\textup{MB}}$, revealing three interesting structures that pervade through the bridge region. First, two distinct filaments peel away from the LMC's western edge, one sits at $\sim 50$ kpc, while the other some 5 kpc farther away. The closer RR Lyrae sequence stretching from the LMC meets half way with the trailing tail of the SMC, emanating from the near end of the dwarf in the direction of the LMC. This appears to be consistent with our findings in Sec.~\ref{sec:smc_shape}, where we saw evidence for (likely unbound) giants lying in front of the SMC being kinematically disrupted towards the larger Cloud. The interpretation of the nature of the second (more distant) part of the bifurcated structure attached to the LMC at $\sim 55$ kpc is more challenging. Are these LMC stars that have been stripped towards the SMC or vice versa? In the right panel, we show a column-normalised version of this representation. The distance gradient as mapped by the RR Lyrae appears to run almost uninterrupted from the east side of the LMC to the far side of the SMC. In this view of the line-of-sight distance distribution, the bifurcation in the LMC side of the old bridge is even more evident (also see the 1D slice shown in the inset). Whilst the exact nature of the old bridge is still unclear, it appears from the RRL that both Clouds are joined by a population of stars over heliocentric distance, whose nature is likely dual, with debris having been stripped from both galaxies. The metallicity distribution of our giants between the Clouds supports this notion; giants with a wide range of metallicities appear to inhabit this region with two main structures apparent. We observe a metal poor continuation of the outer SMC, appearing to extend out to 
$\textup{X}_{\textup{MB}} \sim -15^{\circ}$, consistent with the SMC RRL we observe peeling away from the dwarf's trailing tail. Second to this is the ridge of $\sim -1$ dex that looks to span far into the bridge region, where from the RRL we see such LMC debris exists. We have seen the proper motion distribution of the giants to be turbulent through the bridge region, as stars from both Clouds are being stripped and thrust towards each other in a complex way.  

\section{Conclusions}

We have amassed a sample of Magellanic red giants, drawn from $Gaia$ DR2, demonstrating the ability to predict accurate photometric metallicities for such stars utilising machine learning methods. In doing so, we have produced some of the highest resolution metallicity maps of the Clouds to date. Utilising our metallicity predictions in conjunction with the proper motions of our giants, we produced a chemo-kinematical mapping of the Magellanic system, which shows it to be fraught with intricacy as a consequence of its severe disruption.

We observe negative metallicity gradients on moving outwards through both Clouds. In modelling the LMC as a thin inclined disc, we infer a metallicity gradient of $-0.048 \pm 0.01$ dex/kpc, a value in good agreement other estimates in the literature \citep[see e.g.][]{Cioni_2009, Choudhury_2016}. Centrally, the profiles flatten as this region becomes dominated by the metal-rich bars. Various asymmetries are present in the metallicity profiles when considering different regions of each galaxy. Such features are readily visible in the LMC, where we observe the stellar bar, main spiral arm and diffuse regions of metallicity enhancement in the disc. The most striking example is perhaps the southern spiral arm feature seen in the inner regions, a metal-rich component appearing to emanate from the western end of the bar and wrapping clockwise through the LMC disc. Spatially, this spiral-like feature is coincident with the stellar over-density observed by \citet{Choi_2018_b} and it is very likely that historic LMC-SMC interactions have given rise to this intermediate aged stellar structure. We further observe a northern arc-like region of relative metal enhancement, coincident with the stellar feature observed by \citet{Besla_2016} in their optical images of the LMC periphery. In their simulations of the Magellanic system, they concluded it is the repeated close encounters with the SMC that primarily seeds such one-sided stellar structures in the northern portion of the LMC, with the tidal field of the SMC alone able to induce one-armed spirals within the larger Cloud. 

On slicing the Clouds by metallicity, we observe global evolution in the morphology of the two galaxies. Naturally, the majority of outer substructure is observed in the more metal-poor giants, with those that are metal-richer residing much more centrally in both instances. The outer northern and southern thin spiral-like arms in the LMC are again observed to rotate with the Cloud, lagging in their orbits, as was first noted by \citet{Clouds}. The metal poor SMC giants located nearest to the LMC are hot in their motions, with high dispersion observed at the interface between the two dwarfs, a region where we naturally expect a mixture of stars originating from both the LMC and SMC. Further to this, in Fig.~\ref{fig:pmL}, we observe the eastern portion of the SMC to bear increased motion in the direction of its larger counterpart. On studying the iso-density contours of the SMC, we observe them to display the S-shape signature characteristic of tidal disruption, a feature that appears to persist at all metallicities. We probe the SMC disruption further through a simple consideration of the proper motions of our giants. Utilising each star's relative position in the CMD as a proxy for heliocentric distance, we ascertain that it is those giants likely lying closer to us that exhibit such fervent motion towards the larger Cloud. Our interpretation is that we are observing a tidal tail trailing the SMC, projected along our line of sight, that has been stripped away from the dwarf's core by the LMC. We trace this disruption down to the inner $\sim 2^{\circ}-3^{\circ}$ of the SMC, consistent with the foreground red clump population observed by \citet{Subramanian_2017} which they tracked down to the inner $\sim 2$ kpc (see \citet{Nidever_2013} also). In essence, this stripped population constitutes the start of the old stellar bridge emanating away from the SMC towards the LMC. We do not observe any leading counterpart, and it may simply be that we do not possess enough distant giants in our sample to do so, as such debris is expected to lie $\sim 20$ kpc directly behind the SMC \citep{Diaz_2012}. 

Finally, we consider giants lying in between the Clouds in the Magellanic Bridge region. We, for the first time, trace the metallicity distribution through the region and find its nature to be dual; we observe a metal-poor component of $\feh \sim -1.5$ dex lying just beyond the eastern edge of the SMC, with a continuous metallicity ridge of approximately -1 dex appearing to span much of the distance between the two Clouds. We further consider the $Gaia$ DR2 RR Lyrae around this region, with the added benefit that we can easily estimate the heliocentric distances for such stars. Two thin appendages appear to link the Clouds, lying at distances of $\sim 50$ and 55 kpc. The SMC morphology appears cigar-like, with stellar debris appearing to emanate from the near-end of the dwarf, warping towards the LMC. It also appears that a population of LMC stars extend into the bridge region, indicating that the old stellar populations inhabiting the bridge are a mixture of stripped stars from both the LMC and the SMC.

\section*{Acknowledgements}
JG thanks the Science and Technology Facilities Council of the United Kingdom for financial support.

\section*{Data Availability}
We provide the sample of 226\,119 Giants described in Sec.~\ref{sec:data}, hosted at \url{https://zenodo.org/record/4077356}. The catalogue contains $Gaia's$ $\texttt{source\_id}$, photometry and proper motion measurements alongside 2MASS and WISE photometry. We provide extinction values corrected in the manner explained in Sec.~\ref{sec:data}, as well as our metallicity predictions with associated errors according to Table.~\ref{table:errors}.




\bibliographystyle{mnras}
\bibliography{bibliography} 








\bsp	
\label{lastpage}
\end{document}